\documentclass[a4paper,11pt]{article}
\pdfoutput=1
\usepackage{jinstpub} 
\usepackage{graphicx}
\usepackage{subcaption}
\usepackage{float}
\usepackage{multirow}
\usepackage[linesnumbered,ruled,vlined]{algorithm2e}
\usepackage{booktabs}
\usepackage{lineno}
\usepackage{soul}
\usepackage{comment}

\title{CONCERTO: Optimization of readout electronics}

\author[a,1]{M. Abdkrimi,\note{Corresponding author}}
\author[b]{O. Rossetto,}
\author[b]{O. Bourrion,}
\author[b]{C. Hoarau,}
\author[b]{C. Vescovi}

\affiliation[a] {Univ. Grenoble Alpes, CNRS, Grenoble INP, Institut Néel, Grenoble, 38000, France }

\affiliation[b] {Univ. Grenoble Alpes, CNRS, Grenoble INP, LPSC-IN2P3, Grenoble, 38000, France }

\emailAdd{mounir.abdkrimi@gmail.com}

\abstract{

The CONCERTO millimeter-wave spectral-imaging instrument was deployed on the Atacama Pathfinder EXperiment (APEX), where it acquired science data between April 2021 and May 2023. The instrument features two focal-plane arrays, each composed of 2400 Microwave Kinetic Inductance Detectors (MKIDs). Each array is divided into six feedlines containing 400 MKIDs each, with each feedline read out by a dedicated FPGA-based board, KID\_READOUT.
The next-generation instrument aims to double the detector count per feedline, increasing it from 400 to 800 MKIDs. Achieving this requires a substantial scaling of the readout architecture and poses two key challenges for KID\_READOUT: maintaining readout signal integrity and constraining firmware resource usage, as a direct upscaling of the existing design would exceed the available FPGA capacity.
To overcome these limitations, we developed a Python-based, cycle- and bit-accurate digital twin of the full FPGA digital signal processing chain. This model enabled a detailed investigation of internal signal behavior and provided quantitative guidance for firmware optimization. 
Leveraging these insights, we identified the source of two spurs present in CONCERTO data and significantly reduced their amplitudes. At the same time, we achieved substantial reductions in firmware resource usage—39.0\%pt in LUTs, 20.3\%pt in flip-flops, and 28.98\%pt in DSP slices—without degrading readout performance.
The resulting architecture supports more than 800 MKIDs per feedline on the same hardware platform while preserving readout signal quality, offering a scalable and resource-efficient solution for future high-resolution millimeter-wave astronomical instruments.

}


\begin{document}
\maketitle
\flushbottom

\section{Introduction}

The CONCERTO instrument employs arrays of Microwave Kinetic Inductance Detectors (MKIDs) that form a millimeter-wave camera designed to investigate astrophysical phenomena such as the cosmic microwave background and the formation and evolution of stars~\cite{Day2003,klutsch2003modelisation,ward2007protostars}. 
MKIDs are superconducting radio-frequency resonators whose intrinsic superconducting properties yield high quality factors, enabling frequency multiplexing and the simultaneous readout of hundreds of detectors along a single transmission line.

At the Laboratoire de Physique Subatomique et de Cosmologie (LPSC), twelve dedicated FPGA-based readout electronics boards, named KID\_READOUT, were designed and implemented for the CONCERTO instrument to read out twelve feedlines, each coupled to 400~resonators whose self-resonant frequencies are distributed over a 1\,GHz bandwidth.

Next-generation cameras aim to double their resolution by increasing the number of MKIDs per feedline from 400 to 800. This up-scaling places significant demands on the readout electronics, which must preserve signal integrity while also using FPGA resources efficiently, as up-scaling the existing firmware exceeds the capacity of the KID\_READOUT's FPGA.

In the absence of a detailed model of the digital part of the readout electronics, it is difficult to determine where resource usage can be reduced without degrading the SNR of the readout signal, given the complexity of the DSP chain.
This challenge motivated the development of a bit- and cycle-accurate digital twin model of the current DSP chain implemented in KID\_READOUT's FPGA, providing a comprehensive understanding of its behavior and guiding both the analysis and implementation of optimizations.

This paper describes three successive optimizations of the KID\_READOUT digital chain, from board-level measurements to full instrument validation.

\section{Overview of MKIDs array instrumentation chain}\label{sec:mkids-instrumentation}

\begin{figure}[H]
    \centering
    \includegraphics[width=0.9\textwidth,height=0.32\textwidth]{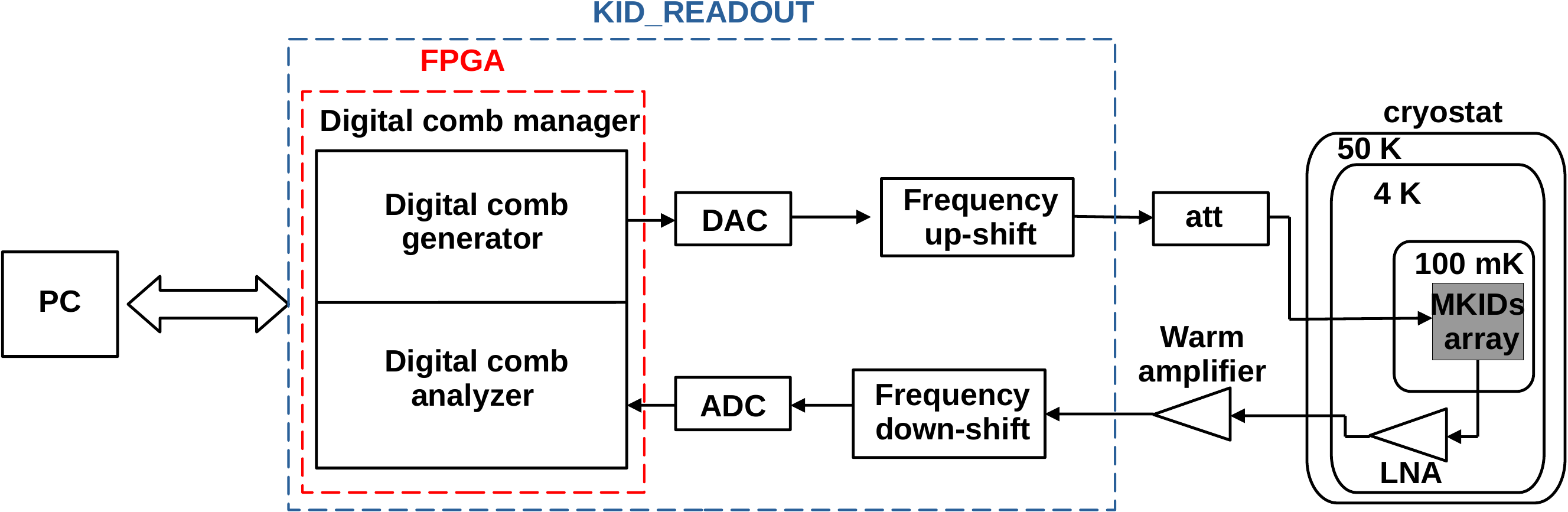}
    \caption{Overview of the instrumentation chain.}
    \label{fig: instrum}
\end{figure}

When a pixel absorbs a photon, the resonance frequency of its corresponding MKID shifts. 
To measure the energy absorbed by each pixel, the FPGA implements a digital comb generator (see Fig.~\ref{fig: instrum}) that produces a synthesized frequency comb consisting of 400~excitation sinusoids (commonly referred to as “tones” in the MKID literature) spanning a large bandwidth of 1\,GHz.

\begin{figure}[H]
    \centering
    \includegraphics[width=1\textwidth]{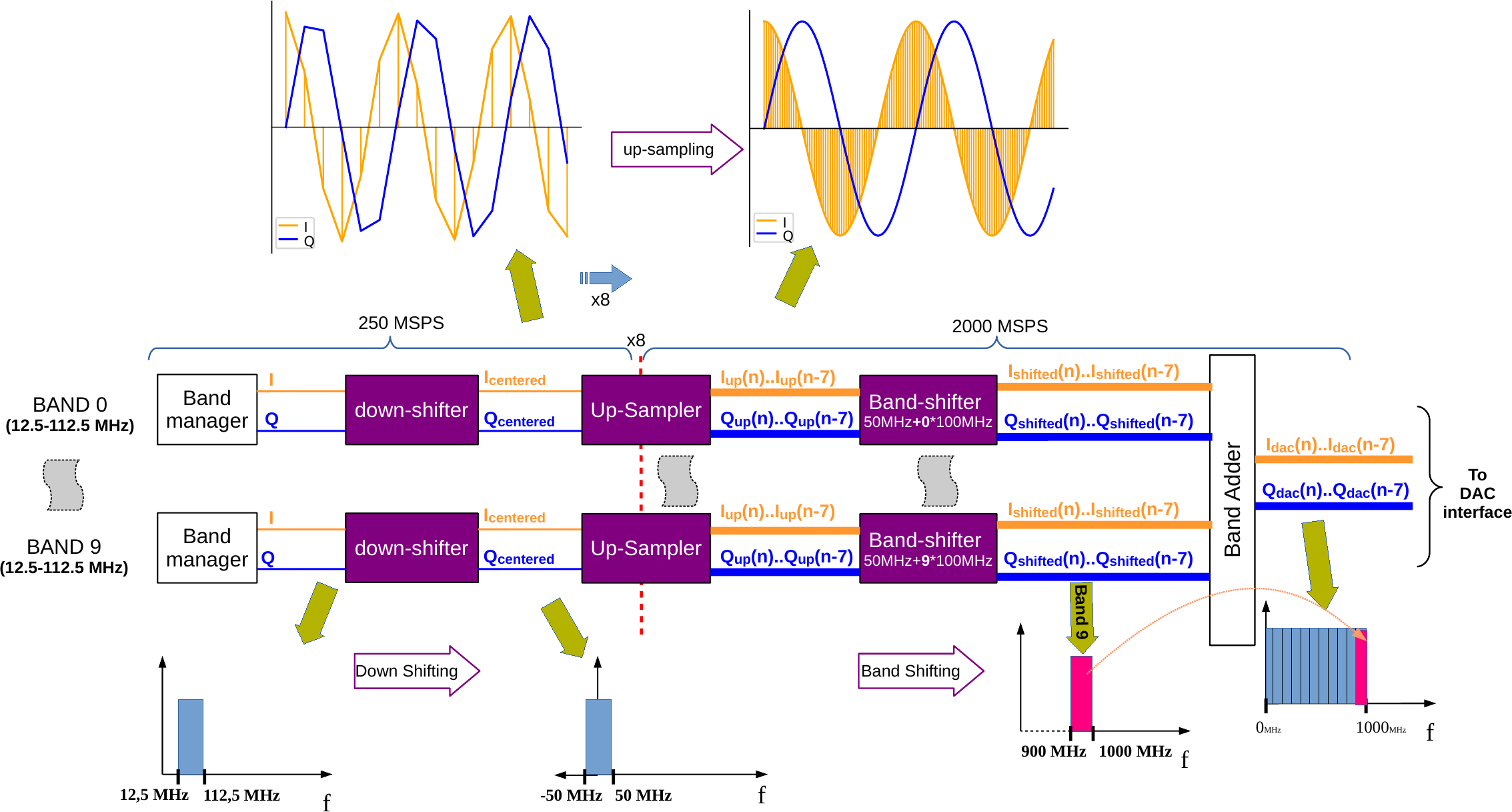}
\caption{Block diagram of the digital frequency comb generator. 
The output spectrum at the end of the chain (sent to the DAC interface) consists of ten sub-bands, each 100\,MHz wide.
Each sub-band is, first, generated by its respective band manager (see its architecture in Figure~\ref{fig:Bandmanagerarchi}) at 250\,MSPS, then frequency-centered around [-50\,MHz 50\,MHz], upsampled to 2\,GSPS and shifted to its corresponding frequency band by the down-shifter, up-sampler and band shifter, respectively.
The figure shows an example for Band~9. 
Finally, the band adder combines the ten bands into a single in-phase/quadrature signal covering the entire [0–1\,GHz] band.}

    \label{fig:firmwaree}
\end{figure}

\begin{figure}[H]
    \centering
    \includegraphics[width=1\textwidth]{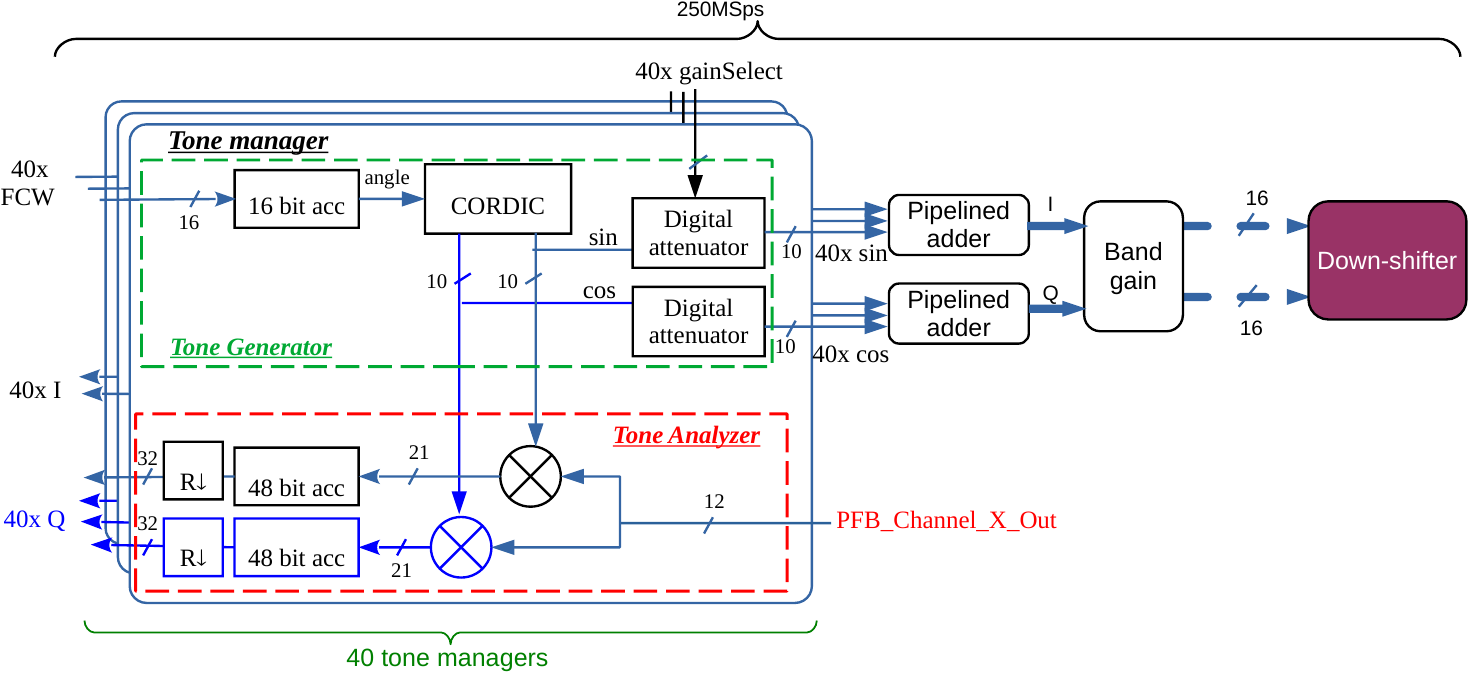}
    \caption{Band Manager architecture.
    It is composed of 40~parallel tones managers, each of them is composed of a tone generator and a tone analyzer.    
    }
    \label{fig:Bandmanagerarchi}
\end{figure}

The excitation waveform is constructed in a hierarchical manner. 
First, ten frequency subbands are generated in parallel by ten band managers, illustrated in Fig.~\ref{fig:firmwaree}, each covering a bandwidth of 100\,MHz and containing 40~tones sampled at 250\,MHz. 
Within the band managers, each tone is produced individually by a dedicated tone generator, illustrated in Fig.~\ref{fig:Bandmanagerarchi} composed of a phase accumulator followed by a COordinate Rotation for DIgital Computer (CORDIC) block, which converts the accumulated phase into its corresponding I/Q components. 
The 40~tones within each subband are then summed to form a 100\,MHz-wide subband waveform.

The ten parallel subband waveforms are subsequently upsampled by a factor of~8, increasing the sampling rate from 250\,MHz to 2000\,MHz, and digitally frequency-up-shifted to their designated spectral locations (i.e., 0–100\,MHz, 100–200\,MHz, …, 900–1000\,MHz) by the band-shifter implemented using a 40-sample sine/cosine lookup table (LUT). Finally, all shifted subbands are summed to generate a single wideband digital waveform spanning 1\,GHz and containing 400~tones.

Each tone is precisely tuned to the resonance frequency of an MKID in its dark (non-illuminated) state. 
The digitally synthesized waveform is converted to analog, up-converted to the RF band of the MKIDs feedline, typically in the range of 1–3\,Ghz, and routed through attenuators at the different temperature stages of the cryostat.

The signal returning from the MKID array is first amplified by a cryogenic low-noise amplifier (LNA) to preserve signal integrity. 
It is then further amplified at room temperature, down-converted to baseband, digitized by an analog-to-digital converter (ADC), and processed by the digital comb analyzer implemented in the FPGA.

At the analyzer input, the 1\,GHz-wide signal containing the 400~tones is first processed by a polyphase filter bank, which channelizes it efficiently into ten 100\,MHz-wide subbands sampled at 250\,MHz, each containing 40~tones. 
Each subband is then routed to 40~parallel dedicated tone analyzers within the tone manager (see Fig.~\ref{fig:Bandmanagerarchi}), where digital down-converters (DDCs) perform subsequent processing. 
For each tone, the corresponding DDC reuses the reference tone generated by the matching CORDIC block in the excitation chain to perform numerical down-conversion to baseband (0\,Hz). 
The signal is then filtered by an averaging filter and downsampled from 250\,MHz to a final sampling rate of $f_s \approx 3814\,\text{Hz}$.

The resulting 400 complex I/Q data streams are transmitted to a PC, where the amplitude, $\sqrt{I^2 + Q^2}$, and phase, $\arctan\left(\frac{Q}{I}\right)$, are computed offline. 
 These quantities provide a direct measure of the energy absorbed by each MKID~\cite{Day2003}.

This paper focuses only on the architectural aspects that have been optimized. 
Detailed technical descriptions of the complete KID\_READOUT DSP chain can be found in~\cite{bourrion2022concerto}.

\section{Optimizations}
\subsection{Methodology}

To optimize the implemented digital processing chain, we adopted a reverse-engineering approach based on the original KID\_READOUT FPGA firmware, focusing specifically on its signal-processing components.
Each Register-Transfer Level (RTL) block was analyzed at both functional and architectural levels, yielding corresponding floating-point Python models that were validated through functional tests and then refined for cycle- and bit-accurate correspondence with the RTL outputs.  

By assembling these validated models and parameterizing each one, we constructed a comprehensive digital twin of the complete DSP readout chain, enabling extensive simulation campaigns to assess performance across various configuration scenarios, identify imperfections, trace their propagation, and evaluate performance–resource trade-offs.

The primary motivation for adopting this approach lies in the efficiency of the Python-based simulation, which produces the required data for our entire DSP chain within roughly one hour, compared to the tens of hours needed for an equivalent VHDL simulation.
Moreover, Python offers convenient built-in tools for data visualization and performance analysis within the same environment.

Finally, the optimized configurations were implemented on the FPGA and validated using KID\_READOUT loopback tests.
The configuration incorporating all proposed optimizations was then deployed, and its performance was thoroughly evaluated and confirmed through instrument-level measurements.

\subsection{Spurs}

Since the KID\_READOUT boards were deployed on CONCERTO, two spurs at fixed frequencies (763\,Hz and 1526\,Hz) have been consistently observed in the I/Q components of all 400~readout channels~\cite{bounmy2022concerto}, as illustrated in Fig.~\ref{fig:setup2_all}. 
Because the complete readout chain involves digital electronics, RF stages, cryogenic components, and the MKID array itself, determining the origin of these spurs remained challenging.

Using the digital twin model of the DSP chain, we reproduced these spurious components in simulation, confirming their digital origin. 
The mechanism arises from a mismatch in periodicity between the excitation path (digital comb generator) and analysis path (digital comb analyzer).
In the excitation chain, individual tones are generated by the tone generator—a $2^{16}$ length phase accumulator and CORDIC unit, as described in Section~\ref{sec:mkids-instrumentation}—yielding a base periodicity of $2^{16}$ samples regardless of tone frequency; 
Subsequent up-sampling by a factor of 8, followed by digital frequency shifting using the 40-sample LUT-based band shifter illustrated in Fig.~\ref{fig:firmwaree}, extends this periodicity to $2^{16} \times 8 \times 5$ samples. A detailed explanation of this extended periodic structure is provided in~\cite{abdkrimi2026spurs}.

When the returning signal enters the analysis chain, the polyphase stage performs complex demodulation, downsampling, and filtering, but some components associated with the extended periodic structure $2^{16} \times 5$ persist. 

The subsequent DDC averaging filter, with length $2^{16}$, does not reject these specific frequencies and only partially attenuates them. 
After further downsampling, this residual five-sample structure aliases into baseband, producing the observed spurs at 763\,Hz ($f_s/5 \approx 3814$\,\text{Hz}/5) and its harmonic at 1526\,Hz in the DDC output.

Using the developed model, we proposed a mitigation strategy that changes the length of both the phase-accumulator and DDC averaging filter from $2^{16}$ to 65520.
 These modifications suppress the extended periodicity, so that in the excitation chain the period becomes \(65520 \times 8\) instead of \(65520 \times 8 \times 5\), and after the polyphase stage only the 65520-periodic components remain, which are then effectively rejected by the 65520-point averaging filter. In a purely digital loopback configuration, this approach completely removes the two spurious tones. 
 A detailed analysis of this phenomenon is discussed in a previous work~\cite{abdkrimi2026spurs}.

\subsubsection{Resource usage}

The proposed solution requires additional FPGA resources compared to the original implementation. 
This increase stems primarily from modifying the phase accumulator behavior: in the original design, it naturally wrapped at $2^{16}$ due to the fixed register width, implicitly performing a modulo-$2^{16}$ operation at no extra cost. 
In contrast, using a modulus of 65\,520---a non-power-of-two constant---necessitates explicit arithmetic logic, thereby increasing resource utilization.

Table~\ref{tab:resource_usage_65520} summarizes the FPGA resource utilization comparison between the complete readout firmware implementations on the Xilinx XCKU060FFVA1156-2 FPGA.

For reference, the total available FPGA resources are as follows: LUTs: 331 680, Flip-Flops (FFs): 663 360, DSPs: 2760.
 
The 65520-based version increases LUT usage by 10758 LUT (from 215228 to 225986),  a 3.24\%pt growth (from 64.89\% to 68.13\%), and FF usage increases by 26,520 FFs (from 346539 to 346759), a 0.03\%pt growth (from 52.24\% to 52.27\%)—while DSP usage remains unchanged.

\begin{table}[h]
\centering
\caption{FPGA resource utilization for the original and new firmware.}
\small
\begin{tabular}{|p{2.5cm}|p{1.5cm}|p{1.5cm}|p{1.5cm}|p{1.5cm}|p{1.2cm}|p{1.2cm}|}
    \hline
    \textbf{Accumulator/ Averaging Length} & 
    \textbf{CLB LUTs Count} & \textbf{CLB LUTs \% Used} & 
    \textbf{CLB FFs Count} & \textbf{CLB FFs \% Used} & 
    \textbf{DSPs Count} & \textbf{DSPs \% Used} \\
    \hline
    \(2^{16} = 65536\) & 215{,}228  & 64.89\% & 346{,}539 & 52.24\% & 1{,}953 & 70.76\% \\
    \hline
    65{,}520           & 225{,}986  & 68.13\% & 346{,}759 & 52.27\% & 1{,}953 & 70.76\% \\
    \hline
\end{tabular}
\label{tab:resource_usage_65520}
\end{table}

Despite a slight increase in LUT and FF utilization, the modified design remains well within the available resources of the target FPGA, making this trade-off acceptable given the significant spectral improvements discussed in Section~\ref{sec:measurement-setup}.

\subsubsection{Measurement setup}\label{sec:measurement-setup}

\begin{figure}[h]
    \centering
    \includegraphics[width=1\textwidth]{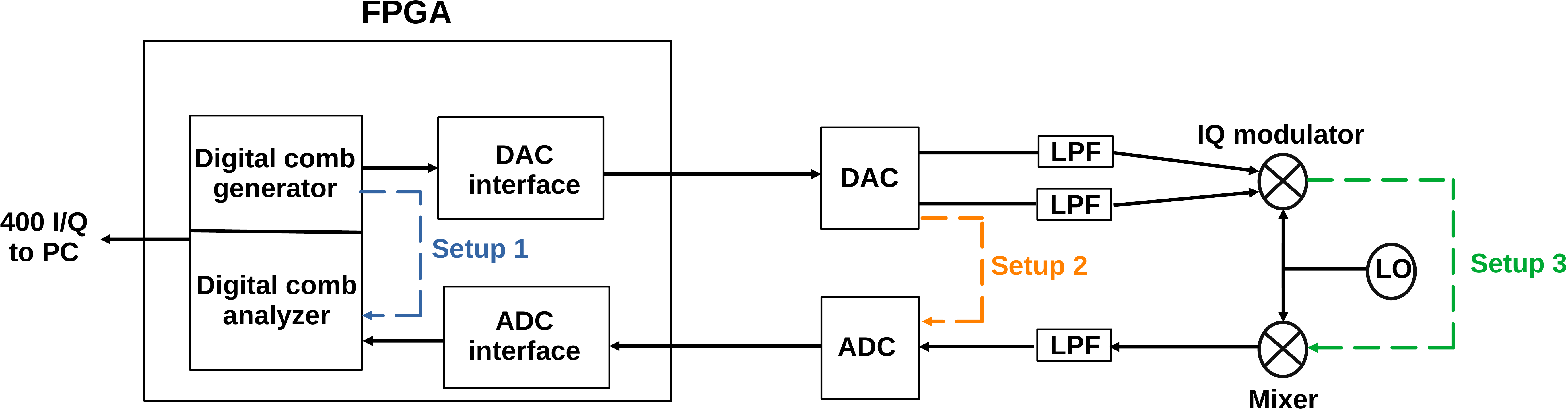}
    \caption{Block diagram of the three measurement setups.  
    Colored paths indicate the feedback routes specific to each setup.}
    \label{fig:setups}
\end{figure}

We conducted a series of acquisition experiments to evaluate both the original and the proposed firmware using Setup~1 (see Fig.~\ref{fig:setups}) in a fully digital configuration. 
The comparison was then extended to Setups~2 and~3, enabling a comprehensive assessment of the spectral performance of the entire electronic readout chain.

Setup~1 consists of a modified firmware configuration where the output of the digital comb generator is directly routed back into the analysis chain, bypassing the DAC, ADC, their respective developed HDL interfaces, and the front-end electronics.
Setup~2 uses the complete firmware, with the signal routed from the DAC output back to the ADC input. 
Finally, Setup~3 includes the complete analog front-end.

For all measurements, 400~I/Q data streams were acquired at system sampling rates over a duration of about 3~minutes.
The acquisition time was chosen to produce the same number of samples per tone—namely, $N = 655{,}360$—ensuring consistent and sufficiently fine frequency resolution for detailed spectral analysis.

\subsubsection{Measurement analysis}  

\textbf{Setup~1}  

\begin{figure}[H]
    \centering

    \begin{subfigure}[t]{0.98\textwidth}
        \centering
        \begin{subfigure}[t]{0.48\textwidth}
            \centering
            \includegraphics[width=\textwidth]{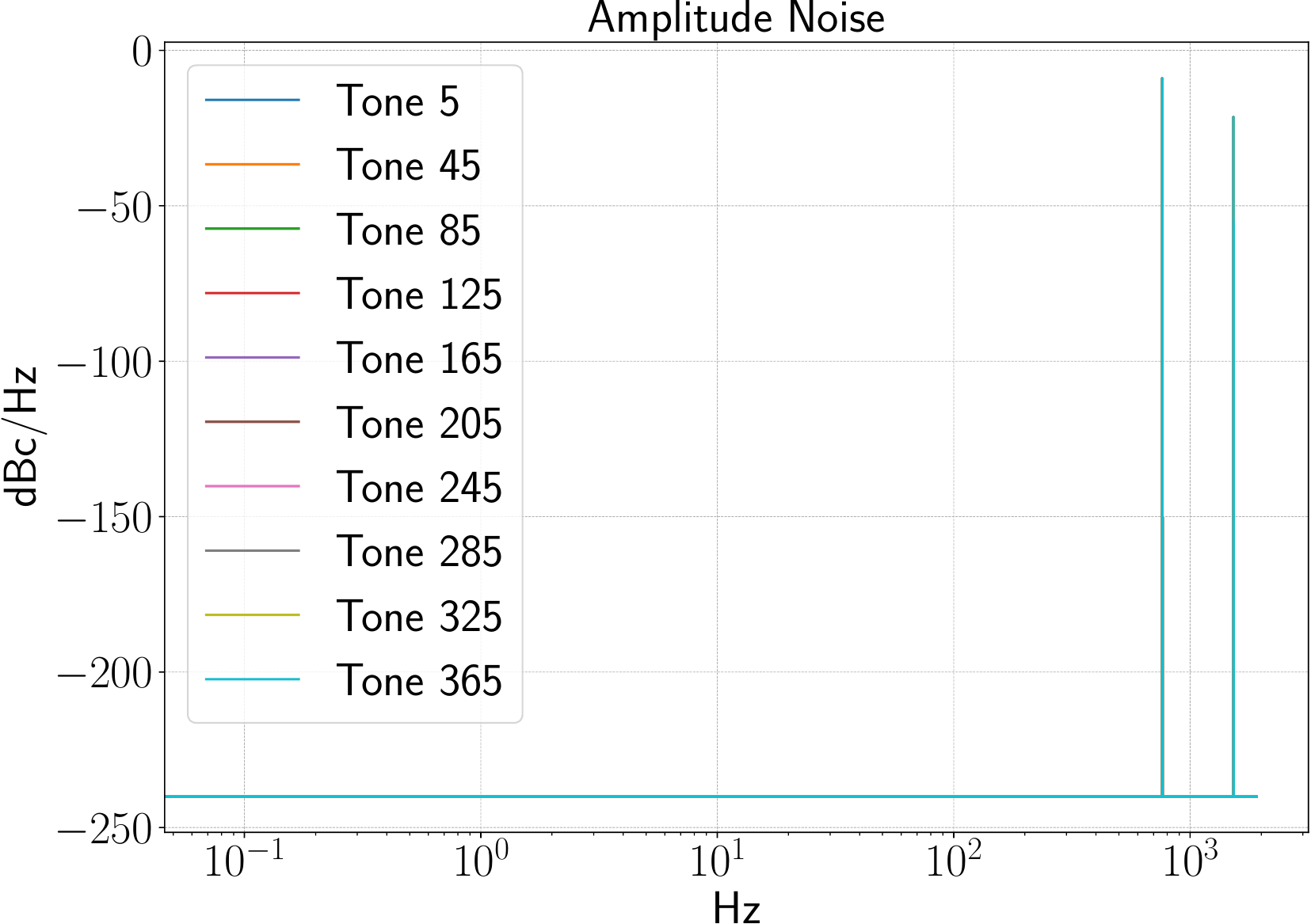}
        \end{subfigure}
        \hfill
        \begin{subfigure}[t]{0.48\textwidth}
            \centering
            \includegraphics[width=\textwidth]{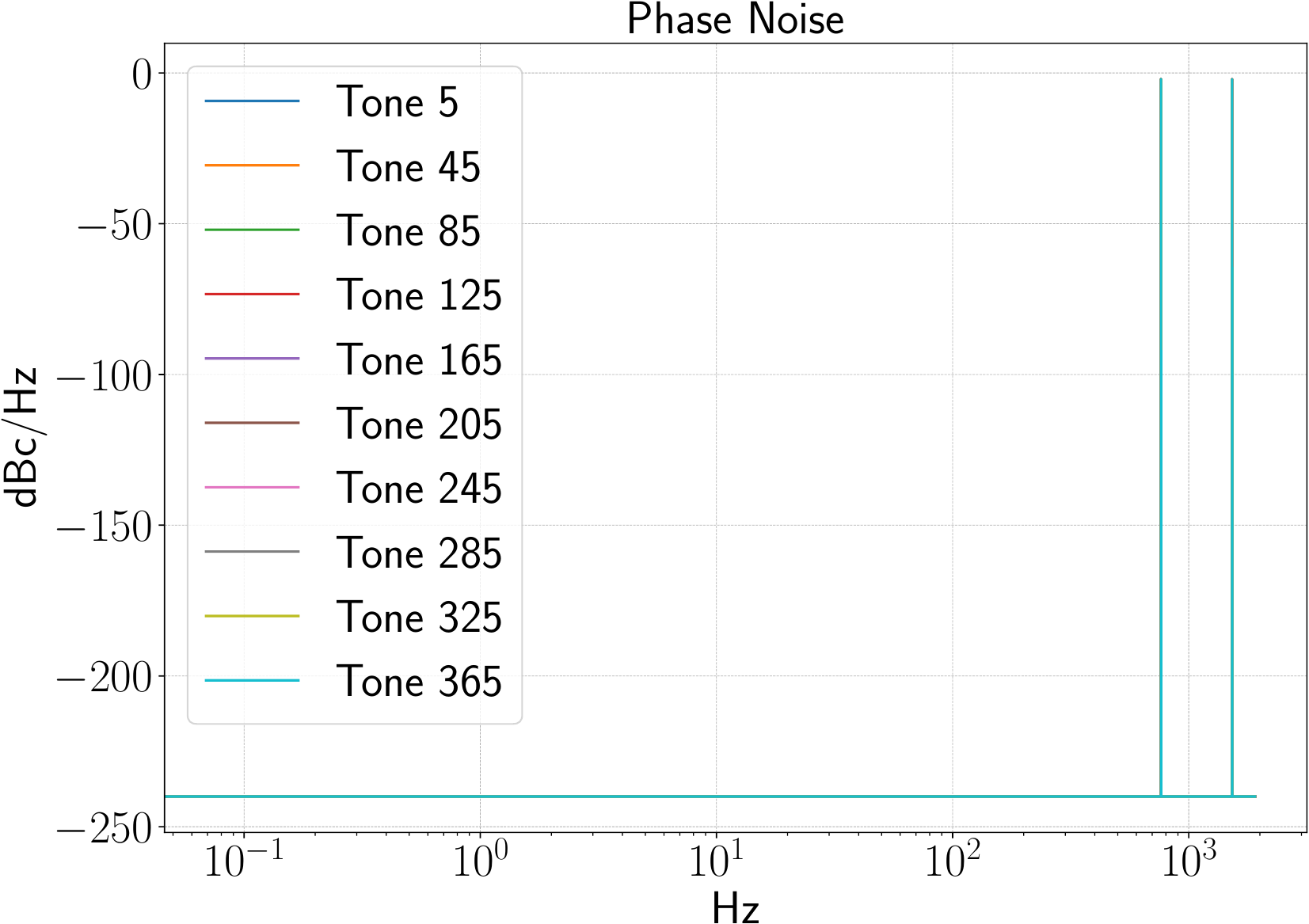}
        \end{subfigure}
        \caption{}
        \label{fig:setup1a}
    \end{subfigure}

    \begin{subfigure}[t]{0.98\textwidth}
        \centering
        \begin{subfigure}[t]{0.48\textwidth}
            \centering
            \includegraphics[width=\textwidth]{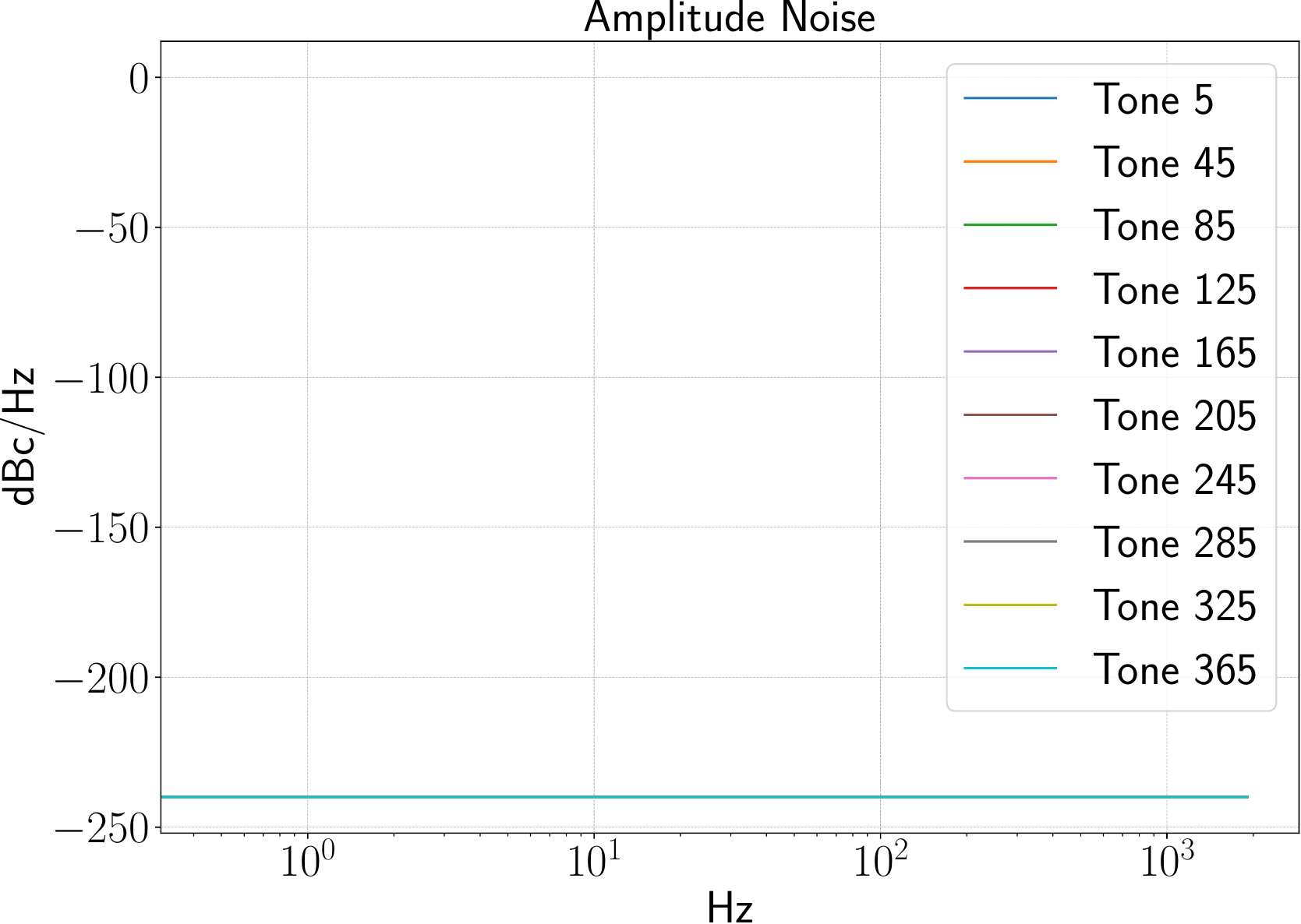}
        \end{subfigure}
        \hfill
        \begin{subfigure}[t]{0.48\textwidth}
            \centering
            \includegraphics[width=\textwidth]{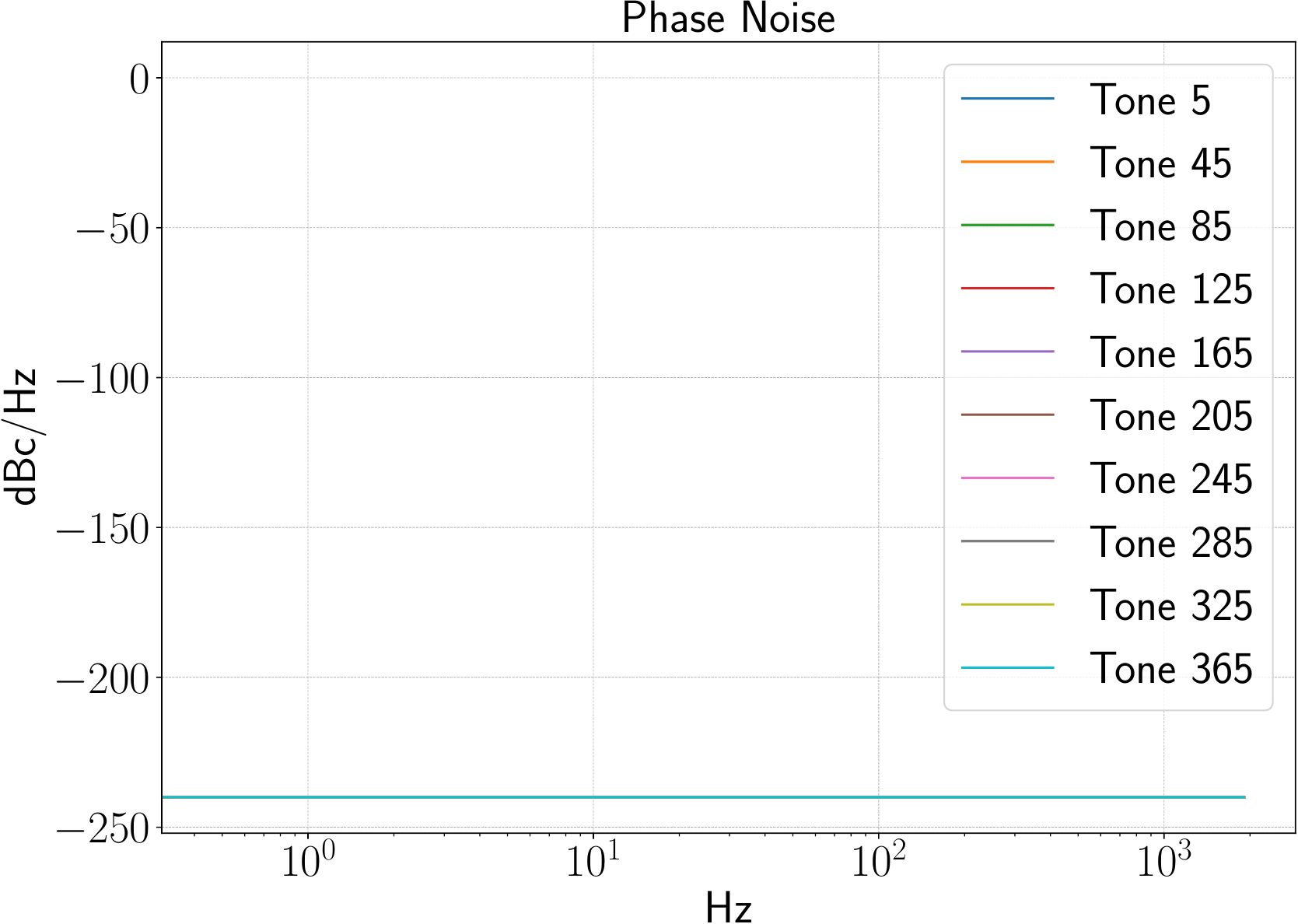}
        \end{subfigure}
        \caption{}
        \label{fig:setup1b}
    \end{subfigure}

\caption{Comparison of measured amplitude and phase noise PSD using Setup 1, between the original firmware (a) and the proposed modified firmware (b). 
A representative tone from each of the 10 frequency bands is shown.}

    \label{fig:setup1_all}
\end{figure}

With the original firmware, the measured spectra exhibit two spurs at 763\,Hz and 1526\,Hz (see Fig.~\ref{fig:setup1_all}\,(a)), confirming through measurement that the phenomenon originates from within the digital processing chain.
As illustrated in Fig.~\ref{fig:setup1_all}\,(b), the amplitude and phase noise spectra show a flat noise floor at $-240$\,dBc/Hz when using the modified firmware, confirming that the DDC averaging filter effectively rejects unwanted spectral components and validating the proposed solution.

\textbf{Setup~2}  
 
\begin{figure}[H]
    \centering

    \begin{subfigure}[t]{0.98\textwidth}
        \centering
        \begin{subfigure}[t]{0.48\textwidth}
            \centering
            \includegraphics[width=\textwidth]{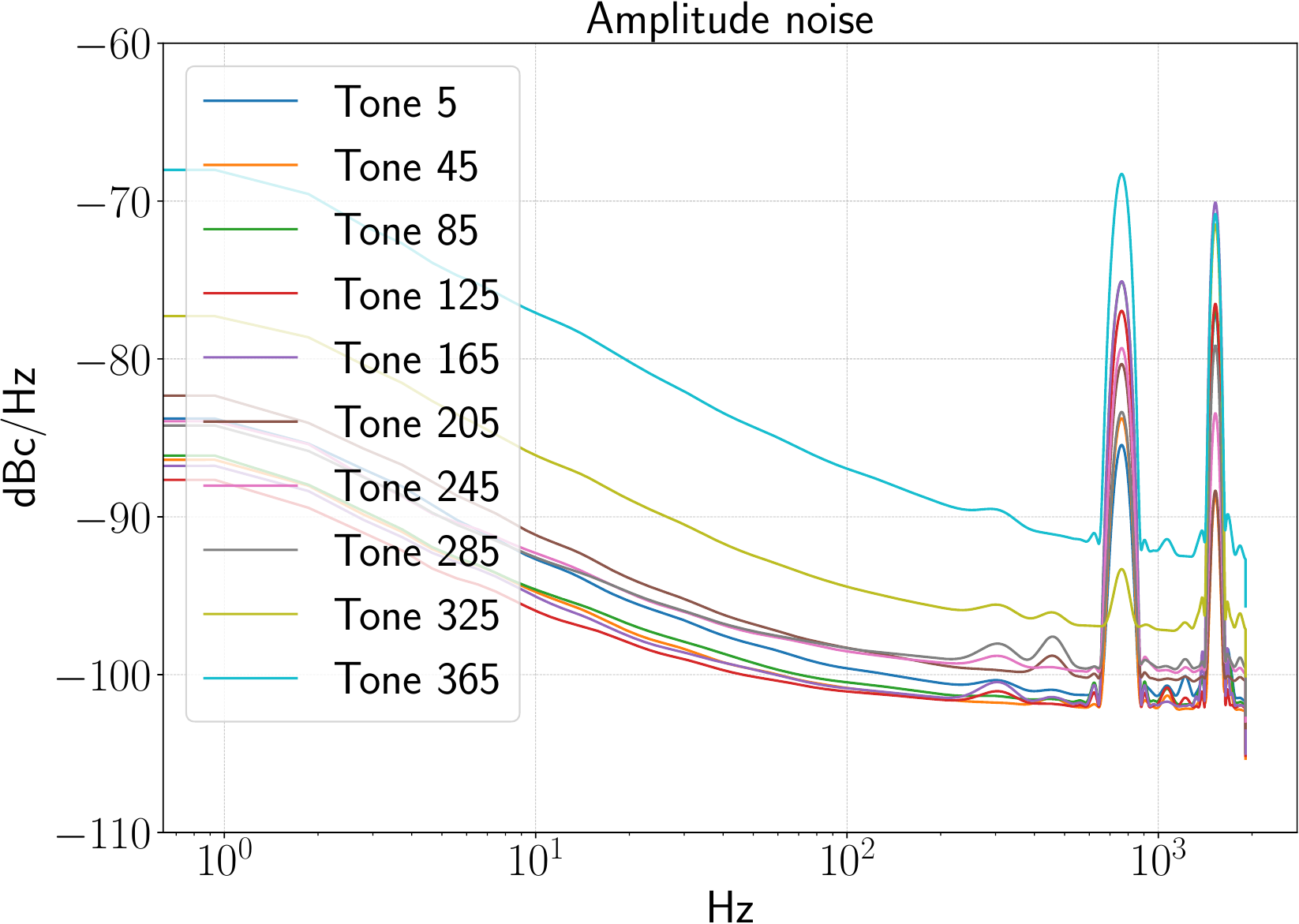}
        \end{subfigure}
        \hfill
        \begin{subfigure}[t]{0.48\textwidth}
            \centering
            \includegraphics[width=\textwidth]{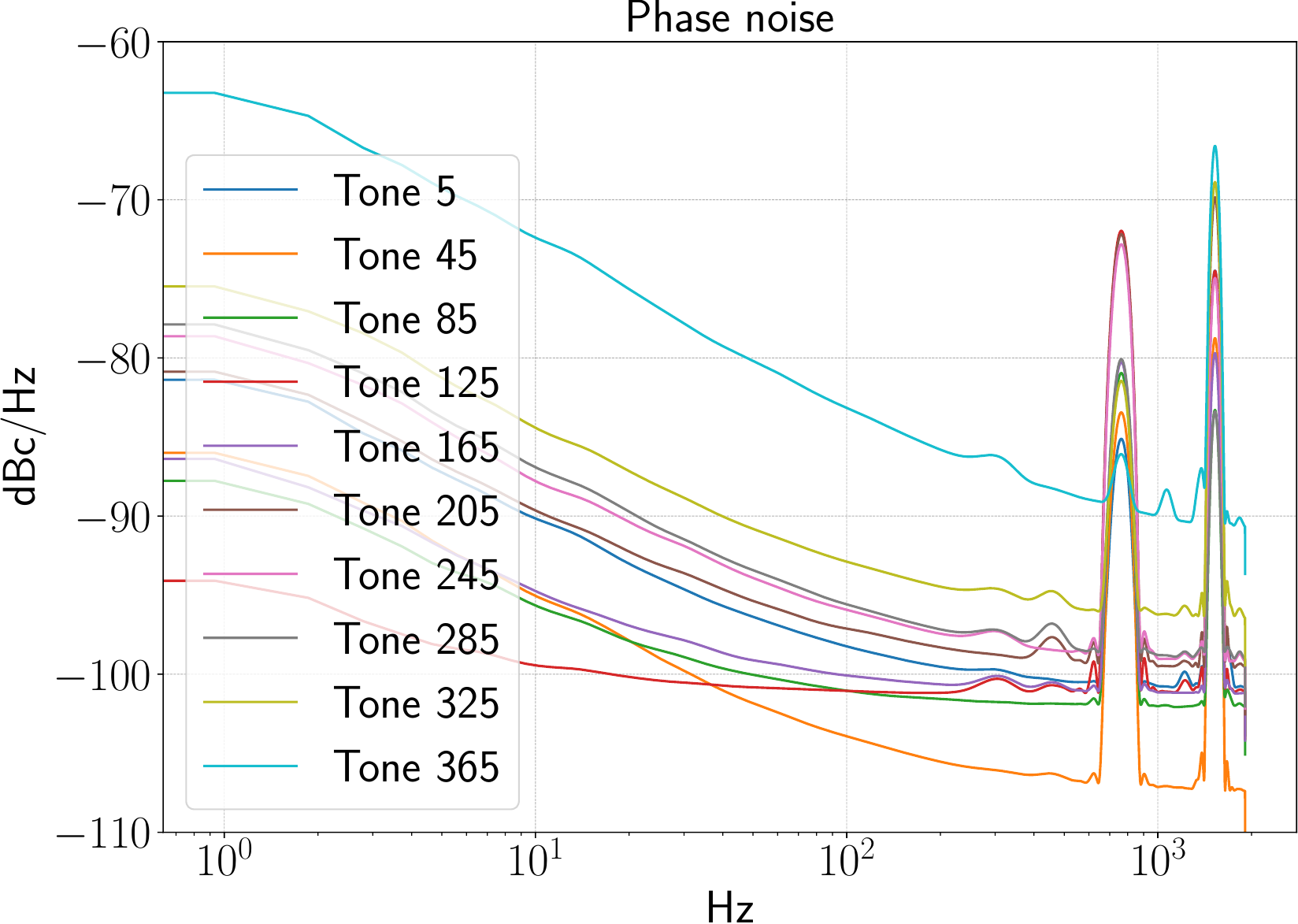}
        \end{subfigure}
        \caption{}
        \label{fig:setup2a}
    \end{subfigure}

    \begin{subfigure}[t]{0.98\textwidth}
        \centering
        \begin{subfigure}[t]{0.48\textwidth}
            \centering
            \includegraphics[width=\textwidth]{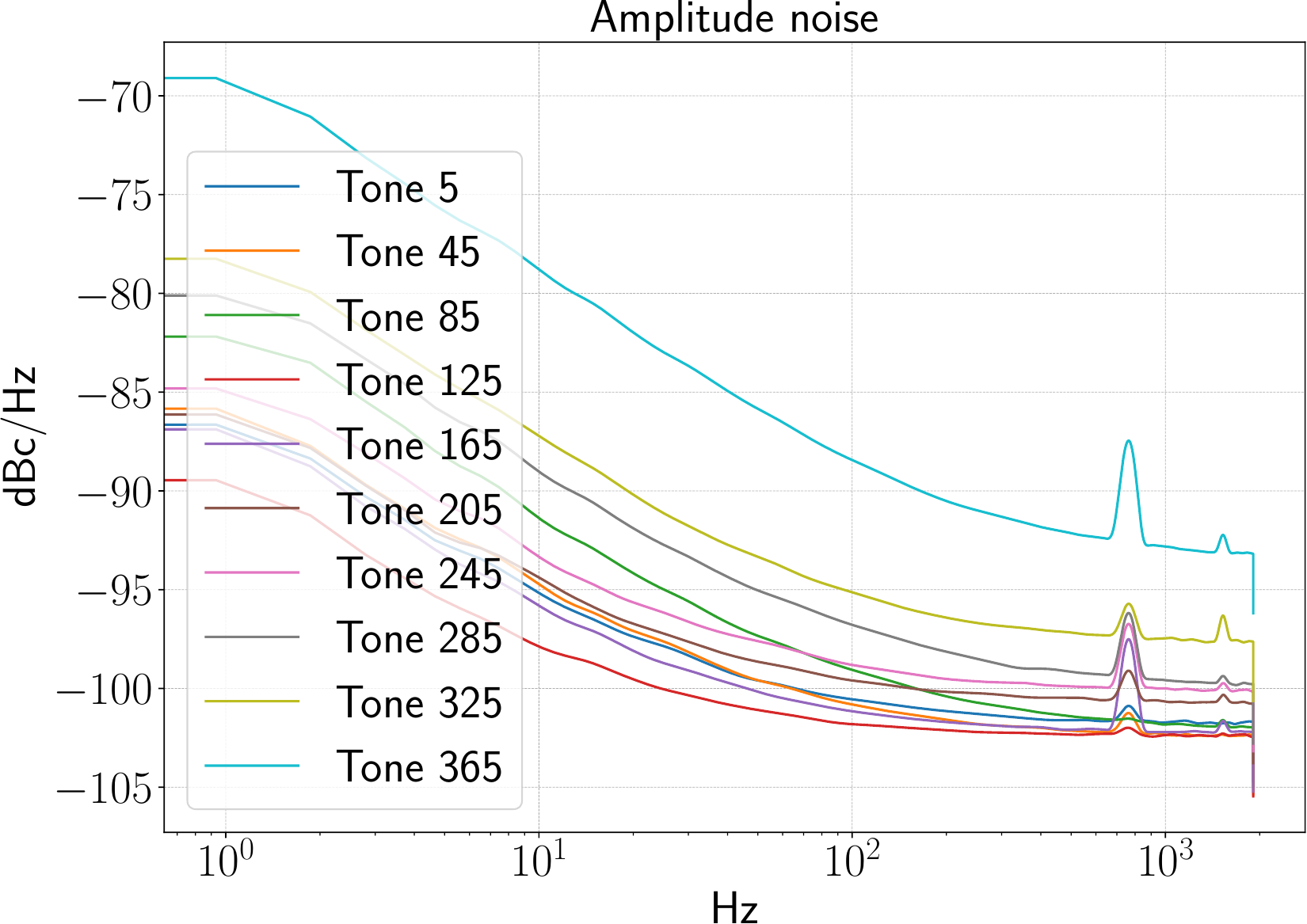}
        \end{subfigure}
        \hfill
        \begin{subfigure}[t]{0.48\textwidth}
            \centering
            \includegraphics[width=\textwidth]{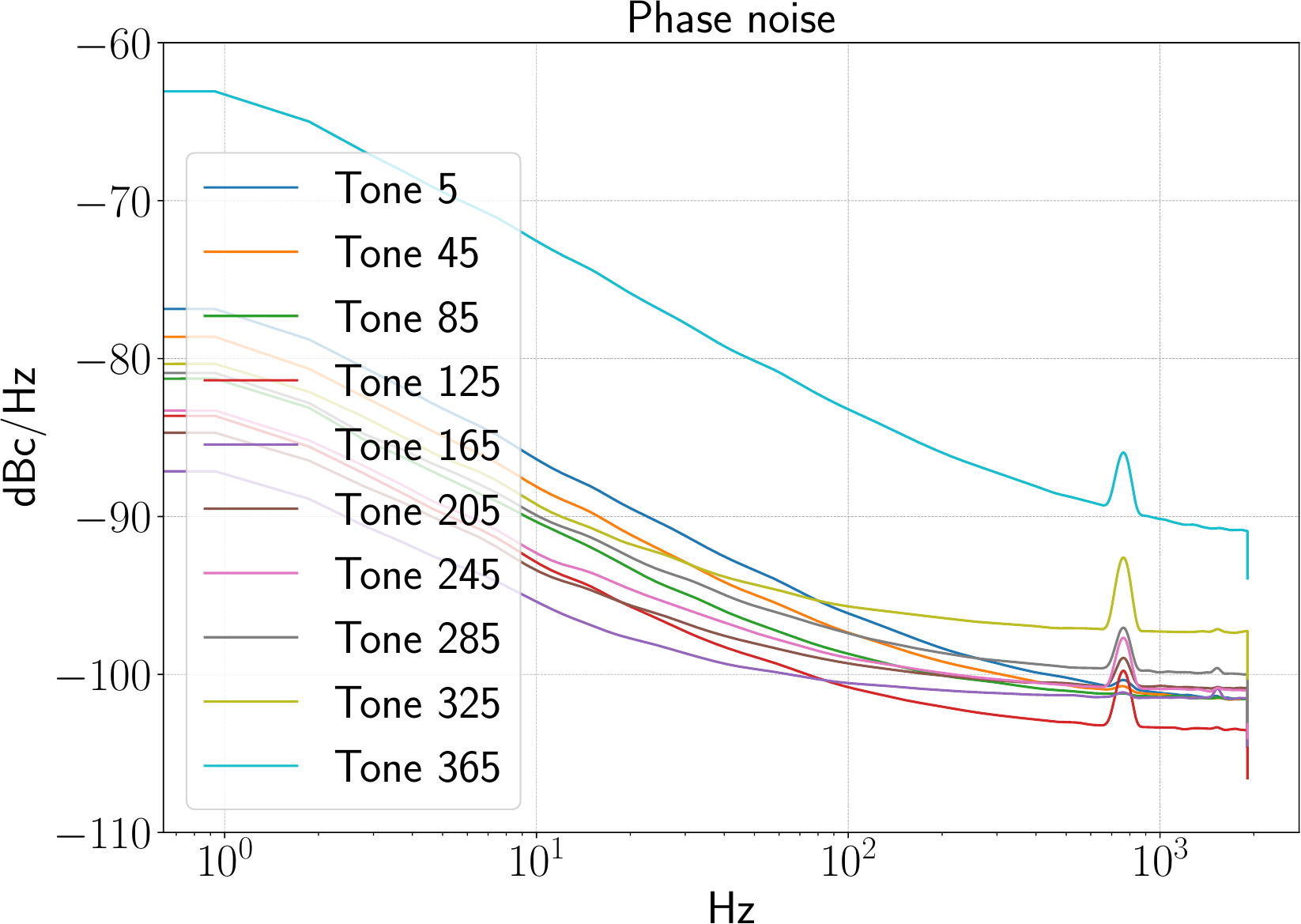}
        \end{subfigure}
        \caption{}
        \label{fig:setup2b}
    \end{subfigure}

\caption{Comparison of measured amplitude and phase noise PSDs using Setup 2, between the original firmware (a) and the proposed modified firmware (b).}

    \label{fig:setup2_all}
\end{figure}

When using the original firmware, spurious tones at 763\,Hz and 1526\,Hz are visible, with peak levels around $-70$\,dB, as illustrated in Fig.~\ref{fig:setup2_all}\,(a).
With the proposed firmware, the 1526\,Hz spur is barely visible, while the 763\,Hz component is attenuated by approximately 25\,dB—reducing its level to below $-95$\,dB for most tones, except in the ninth spectral band (Tone~365).
Due to spectral variability in measurements involving analog components (Setups 2 and 3), Welch’s method is used for spectral estimation because it reduces the variance of the estimated PSD.

\textbf{Setup~3}  

\begin{figure}[h]
    \centering

    \begin{subfigure}[t]{0.98\textwidth}
        \centering
        \begin{subfigure}[t]{0.48\textwidth}
            \centering
            \includegraphics[width=\textwidth]{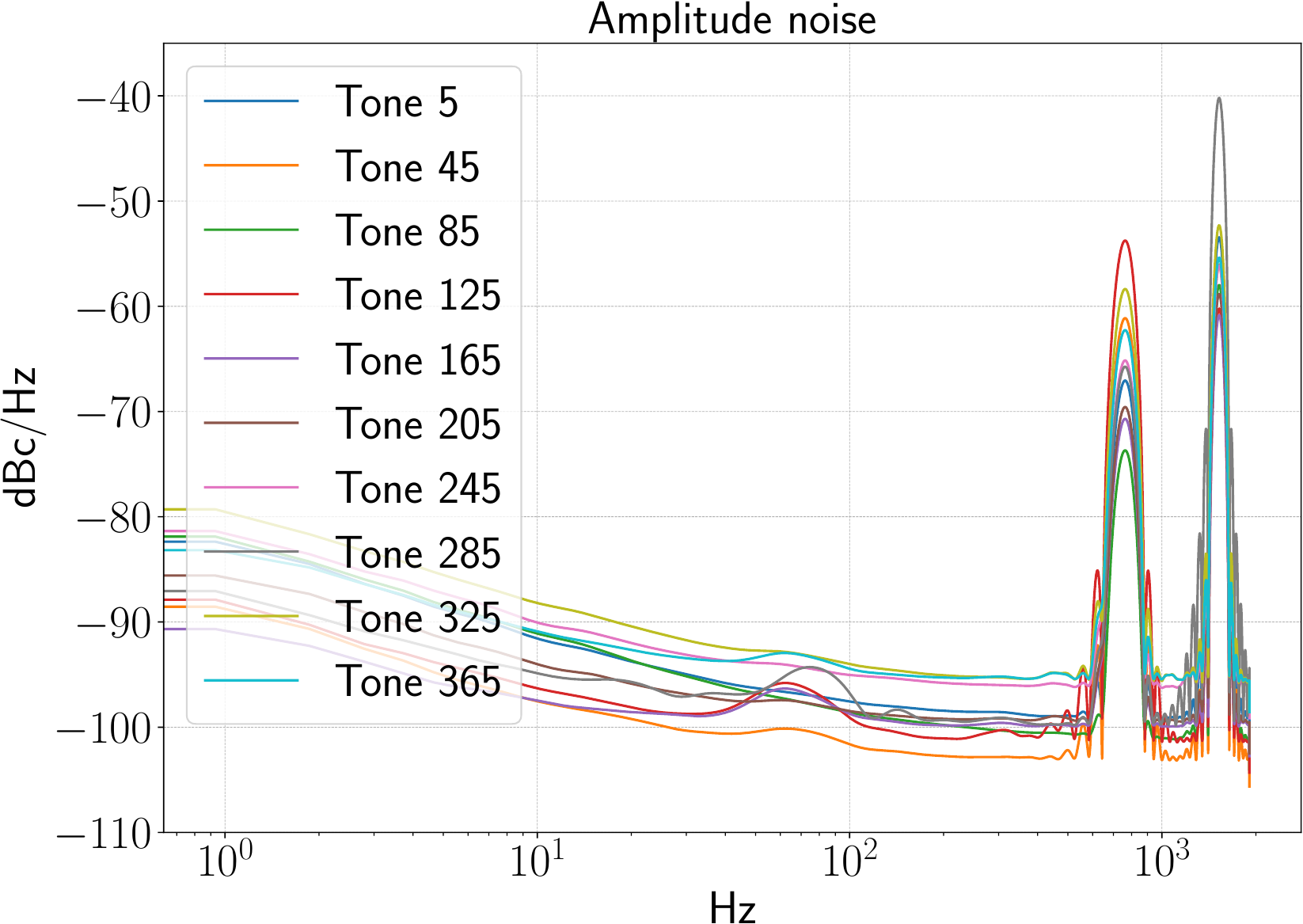}
        \end{subfigure}
        \hfill
        \begin{subfigure}[t]{0.48\textwidth}
            \centering
            \includegraphics[width=\textwidth]{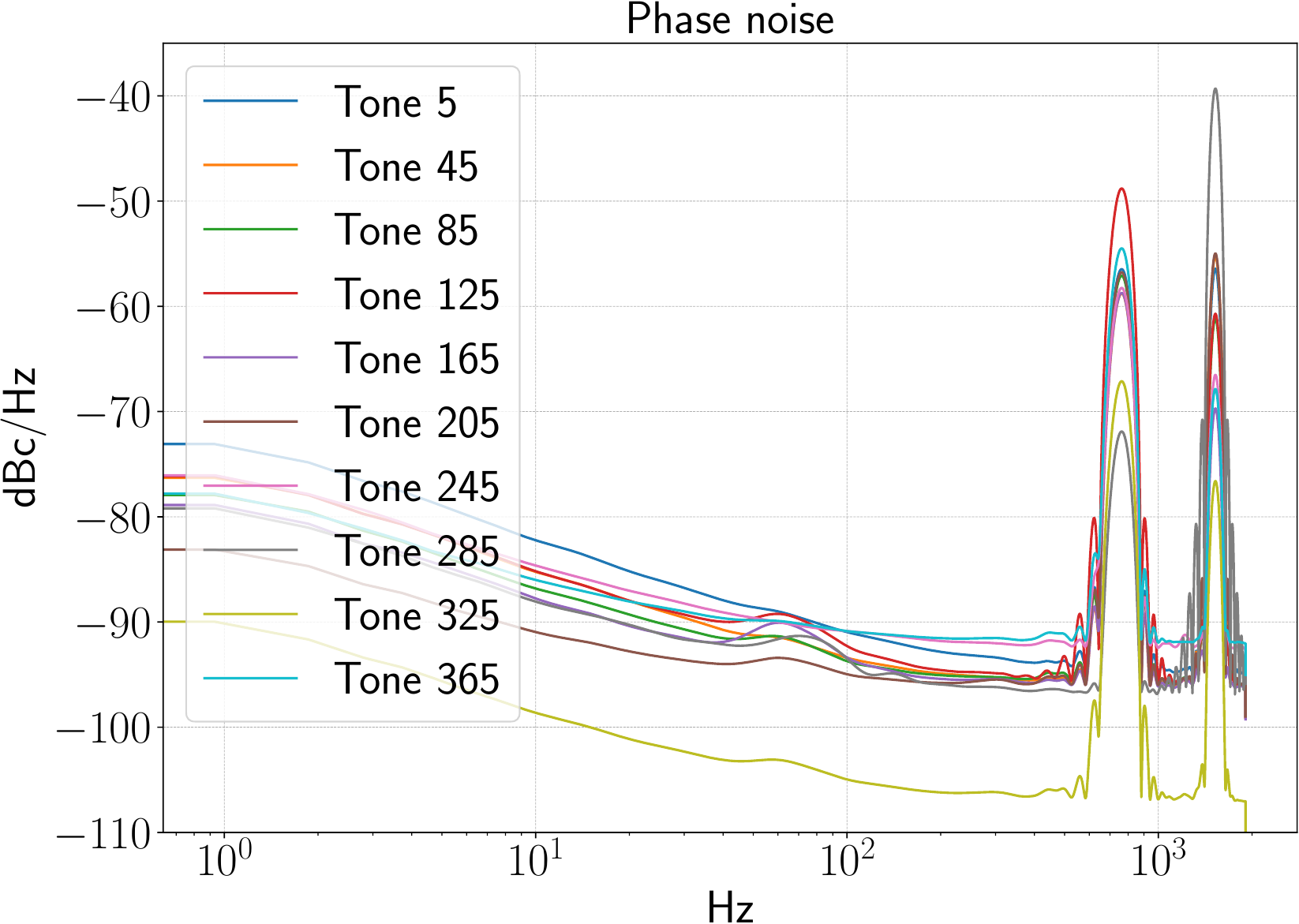}
        \end{subfigure}
        \caption{}
        \label{fig:setup3a}
    \end{subfigure}

    \vspace{0.3cm}

    \begin{subfigure}[t]{0.98\textwidth}
        \centering
        \begin{subfigure}[t]{0.48\textwidth}
            \centering
            \includegraphics[width=\textwidth]{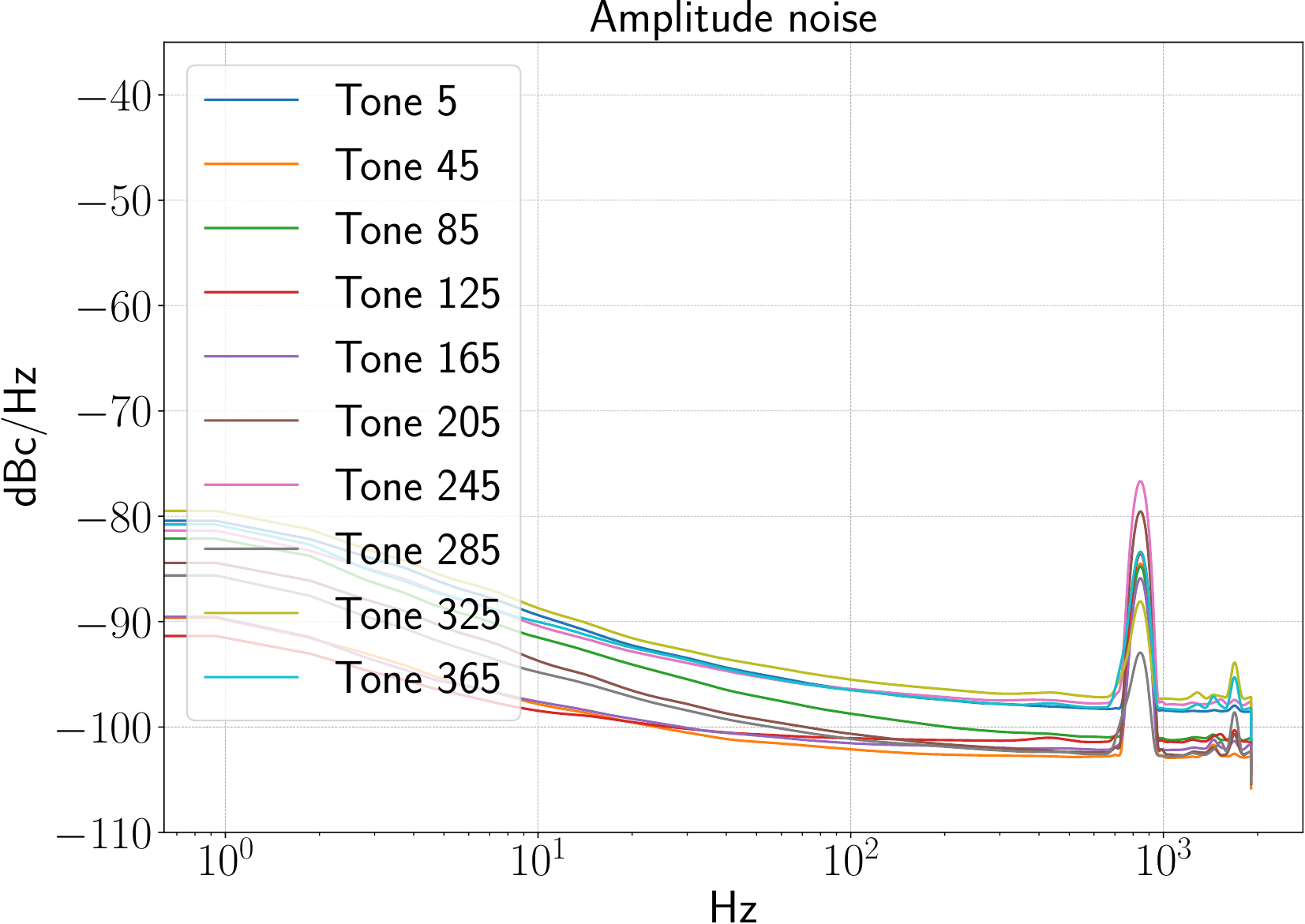}
        \end{subfigure}
        \hfill
        \begin{subfigure}[t]{0.48\textwidth}
            \centering
            \includegraphics[width=\textwidth]{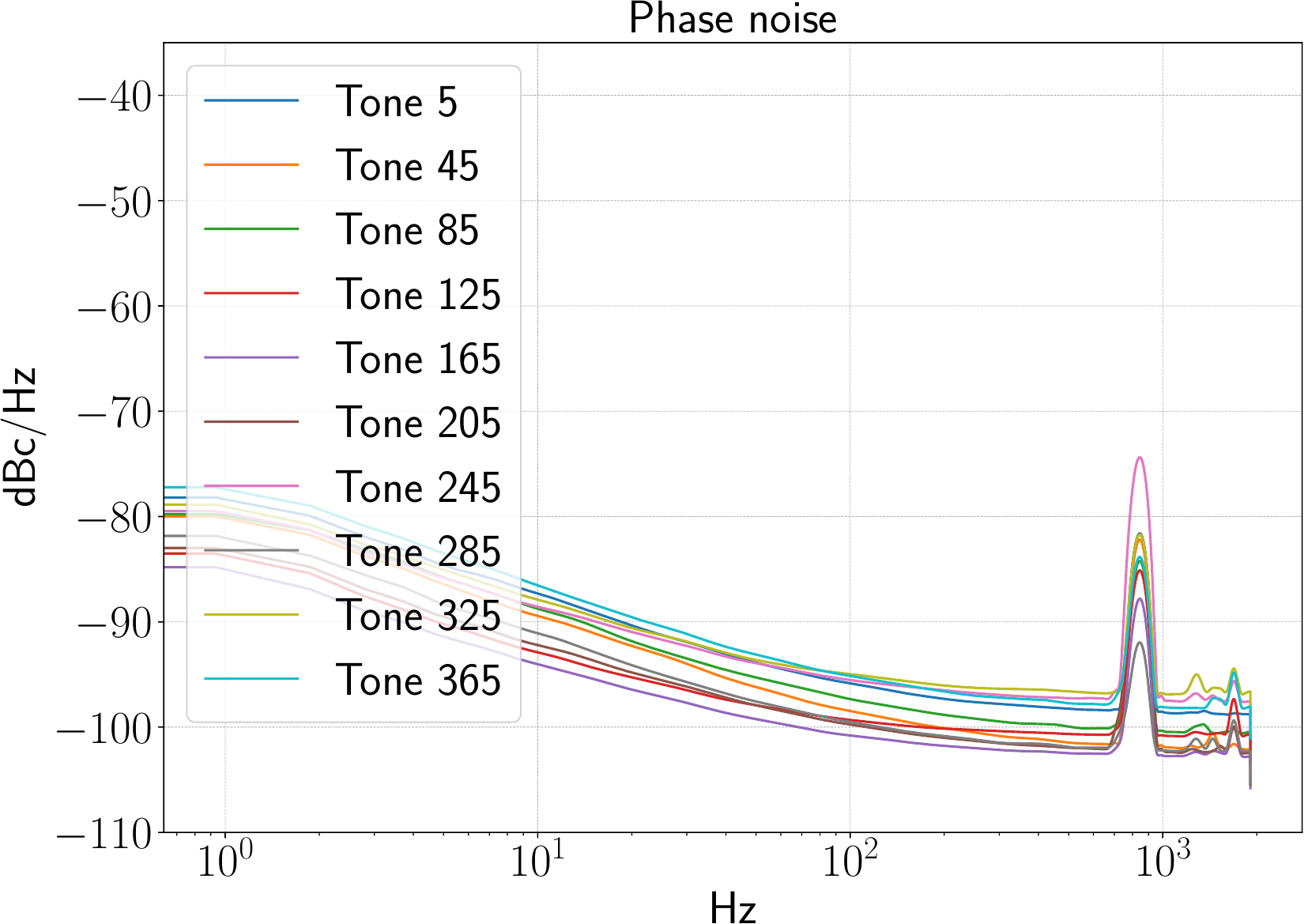}
        \end{subfigure}
        \caption{}
        \label{fig:setup3b}
    \end{subfigure}

\caption{Comparison of measured amplitude and phase noise PSDs using Setup 3, between the original firmware (a) and the proposed modified firmware (b). }

    \label{fig:setup3_all}
\end{figure}

The original firmware exhibits spurs at approximately $-50$\,dB (763\,Hz) and $-40$\,dB (1526\,Hz).  
With the new firmware, these components are significantly attenuated: the 763\,Hz spur is reduced to around $-80$\,dB, and the 1526\,Hz spur falls below $-95$\,dB—corresponding to attenuations of 30\,dB and 45\,dB, respectively (as illustrated in Fig.~\ref{fig:setup3_all})

The observed results show that the transition from Setup~1 to Setup~2 reveals a partial reappearance of the 763\,Hz and 1526\,Hz spurs when using the proposed firmware.
This observation suggests that certain residual periodicity effects—likely related to the $2^{16}$ periodicity—are introduced by the DAC, or the ADC, or one of their VHDL-developed interface components.

Pinpointing the exact source requires further investigation and dedicated development effort.
For example, the DAC interface, implemented within the FPGA, is responsible for formatting the digital excitation comb signal according to the specific protocol required by the DAC.
As a result, probing the output of this interface is not straightforward—the signal is encoded in a hardware-specific format that cannot be directly interpreted or analyzed.
Extracting a usable version would require additional development to implement custom logic capable of capturing and decoding the formatted data stream.
Nevertheless, the implemented mitigation strategy significantly enhances the spectral purity of the readout, reducing artifacts by up to 45\,dB.

\subsection{CORDIC}

The CORDIC algorithm is highly hardware-efficient for implementing trigonometric functions, as it relies exclusively on shift-and-add operations. 
Moreover, its iterative structure makes it particularly well suited for FPGA implementations, since it can be pipelined to deliver one output sample per clock cycle.

Despite this computational efficiency, the current architecture implements 400~parallel CORDIC units, which significantly increases firmware resource utilization. 
The design already consumes 68.13\% of the available LUTs and 52.27\% of the flip-flops.
Such a high utilization limits scalability and makes the migration toward an upgraded architecture with 800~CORDIC units challenging.

\subsubsection{Bit-Width and Iteration Optimization}

In the current implementation, each CORDIC unit comprises 10 pipelined stages (i.e., 10 iterations) and generates 10-bit I (cosine) and Q (sine) components.

\begin{figure}[H] \centering \includegraphics[width=0.75\linewidth]{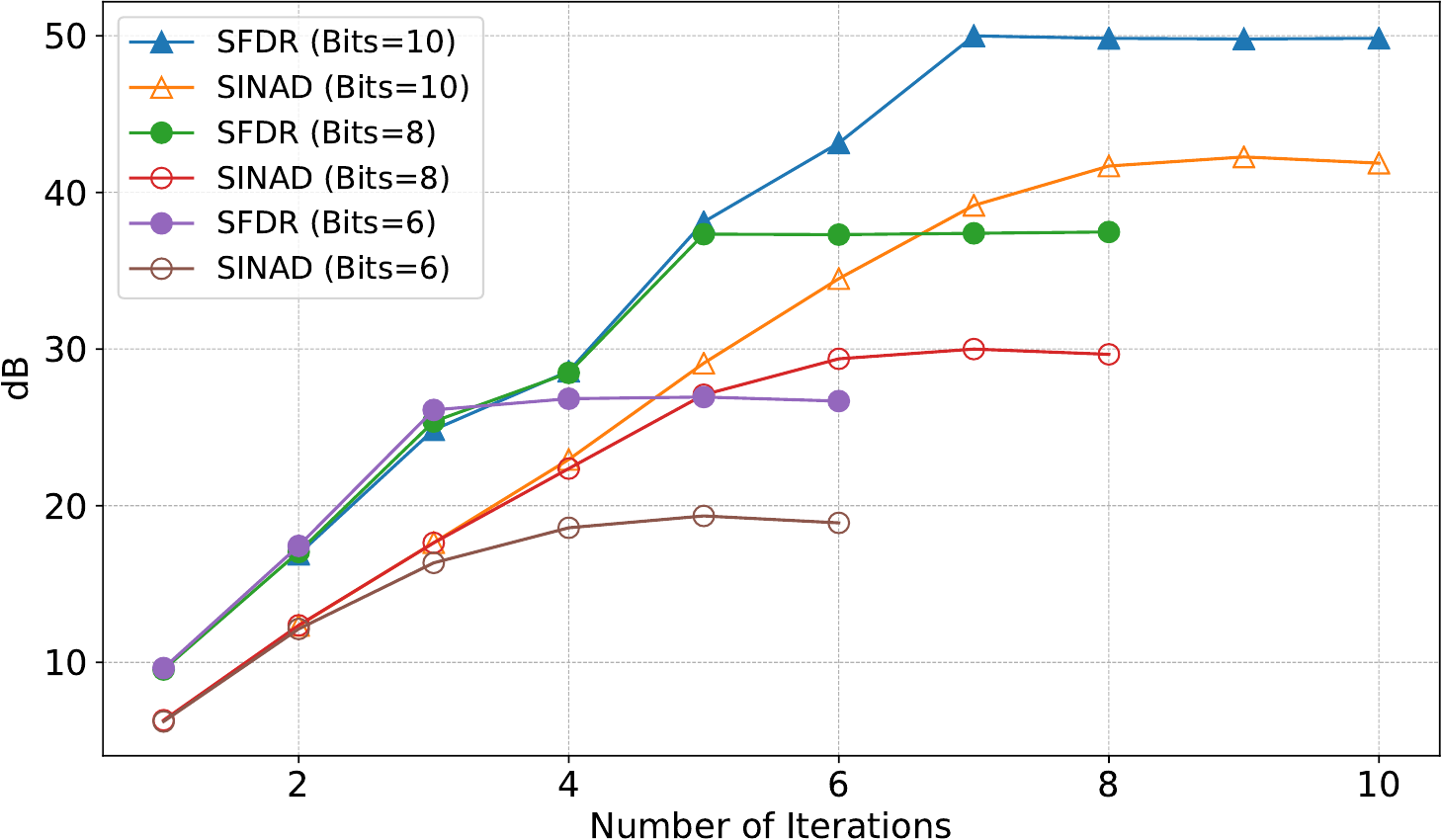} \caption{SINAD and SFDR as functions of number of iterations for different bit widths.} \label{fig:SNR_SFDR} \end{figure}

To determine an optimal configuration that minimizes resource utilization while maintaining acceptable signal quality, we evaluated the impact of reducing both the bit width and the number of CORDIC iterations. The digital frequency tone generator model was used to assess the resulting SINAD and SFDR performance. Fig.~\ref{fig:SNR_SFDR} presents these metrics as functions of the iteration count for various bit widths.

The results show that the SFDR remains unchanged when the number of iterations is reduced by up to three, independently of the bit width configuration. For instance, a 10-bit CORDIC with 10 iterations achieves the same SFDR of 50\,dB as a 10-bit implementation with only 7 iterations. The corresponding SINAD degradation is minimal, with a reduction of approximately 2.5\,dB when three iterations are removed.

This trade-off is highly beneficial. Reducing three iterations removes three pipeline stages per CORDIC unit, resulting in a total reduction of 1200~pipeline stages across the 400~parallel tone generator units. This directly translates into FPGA resource savings, which are quantified in Section~\ref{sec:ressourge_usage_CORDIC}.

Further analysis demonstrated that a 6-bit, 3-iteration CORDIC configuration offers the best compromise between hardware efficiency and signal performance, as detailed in our previous work~\cite{abdkrimi2025cordic}.

\subsubsection{Resource usage}\label{sec:ressourge_usage_CORDIC}

To compare the resource savings with respect to the original solution, we implemented multiple versions of the complete firmware based on the original architecture, which includes the $2^{16}$-sample phase accumulator and averaging filter. 
Each version integrates a different CORDIC configuration, with varying numbers of iterations and bit widths. 

As discussed previously, the configuration with 6-bit precision and 3 iterations provides an optimal trade-off for our system. In practical terms, this configuration yields substantial FPGA resource savings compared to the current implementation (10-bit width, 10 iterations). As summarized in Table~\ref{tab:resource_usage}, LUT usage decreases by 129,403 LUTs (from 215,228 to 85,825), a 39.01\%pt reduction (from 64.89\% to 25.88\%), and FF consumption by 134,058 FFs (from 346,539 to 212,481), a 20.21\%pt reduction (from 52.24\% to 32.03\%).

\begin{table}[h]
    \centering
    \small
    \renewcommand{\arraystretch}{1.3}
    \setlength{\tabcolsep}{6pt}
    \caption{FPGA resource utilization for the entire readout firmware, evaluating different bit width and iteration configurations of the CORDIC implementation.} 
    \begin{tabular}{|c|c|c|c|c|c|}
        \hline
        \textbf{Bits} & \textbf{Iterations} & \multicolumn{2}{c|}{\textbf{CLB LUT}} & \multicolumn{2}{c|}{\textbf{CLB FF}} \\
        \cline{3-6}
        & & \textbf{Count} & \textbf{\% Used} & \textbf{Count} & \textbf{\% Used} \\
        \hline

        \multirow{2}{*}{\centering \textbf{10}} & 10 & 215\,228 & 64.89\% & 346\,539 & 52.24\% \\ 
        \cline{2-6} 
        & 7  & 182\,422 & 55.00\% & 314\,089 & 47.35\% \\ 
        \hline

        \multirow{2}{*}{\centering \textbf{9}} & 9  & 184\,926 & 55.75\% & 312\,641 & 47.13\% \\ 
        \cline{2-6} 
        & 6  & 154\,722 & 46.65\% & 282\,857 & 42.64\% \\ 
        \hline

        \multirow{2}{*}{\centering \textbf{8}} & 8  & 165\,977 & 50.04\% & 288\,345 & 43.47\% \\ 
        \cline{2-6} 
        & 5  & 139\,160 & 41.96\% & 261\,825 & 39.47\% \\ 
        \hline

        \multirow{2}{*}{\centering \textbf{7}} & 7  & 144\,286 & 43.50\% & 265\,273 & 39.99\% \\ 
        \cline{2-6} 
        & 4  & 119\,897 & 36.15\% & 241\,201 & 36.36\% \\ 
        \hline

        \multirow{2}{*}{\centering \textbf{6}} & 6  & 111\,517 & 33.62\% & 236\,897 & 35.71\% \\ 
        \cline{2-6} 
        & 3  & 85\,825 & 25.88\% & 212\,481 & 32.03\% \\ 
        \hline

        \multirow{2}{*}{\centering \textbf{5}} & 5  & 96\,117 & 28.98\% & 222\,801 & 33.59\% \\ 
        \cline{2-6} 
        & 2  & 77\,661 & 23.41\% & 204\,853 & 30.88\% \\ 
        \hline

    \end{tabular}
    \label{tab:resource_usage}
\end{table}

\subsubsection{Measurement analysis}

To confirm that the proposed configuration does not degrade readout signal quality while providing significant resource optimization, measurements were performed using Setup~3 (see Fig.~\ref{fig:setups}), which includes all components of KID\_READOUT.  

We compared the original firmware, which uses a 10-bit, 10~iterations CORDIC, with the optimized version using 6-bit, 3 iterations.   
In both cases, the 65,520 samples solution was used in place of the original \(2^{16}\)-sample configuration to evaluate whether the reduced CORDIC precision preserves the performance benefits of the 65,520 samples optimization.

\begin{figure}[H]
    \centering

    \begin{subfigure}[t]{0.98\textwidth}
        \centering
        \begin{subfigure}[t]{0.48\textwidth}
            \centering
            \includegraphics[width=\textwidth]{figures/setup3_new_amp-cropped.pdf}
        \end{subfigure}
        \hfill
        \begin{subfigure}[t]{0.48\textwidth}
            \centering
            \includegraphics[width=\textwidth]{figures/setup3_new_phase-cropped.pdf}
        \end{subfigure}
        \caption{}
        \label{fig:setup3a}
    \end{subfigure}

    \begin{subfigure}[t]{0.98\textwidth}
        \centering
        \begin{subfigure}[t]{0.48\textwidth}
            \centering
            \includegraphics[width=\textwidth]{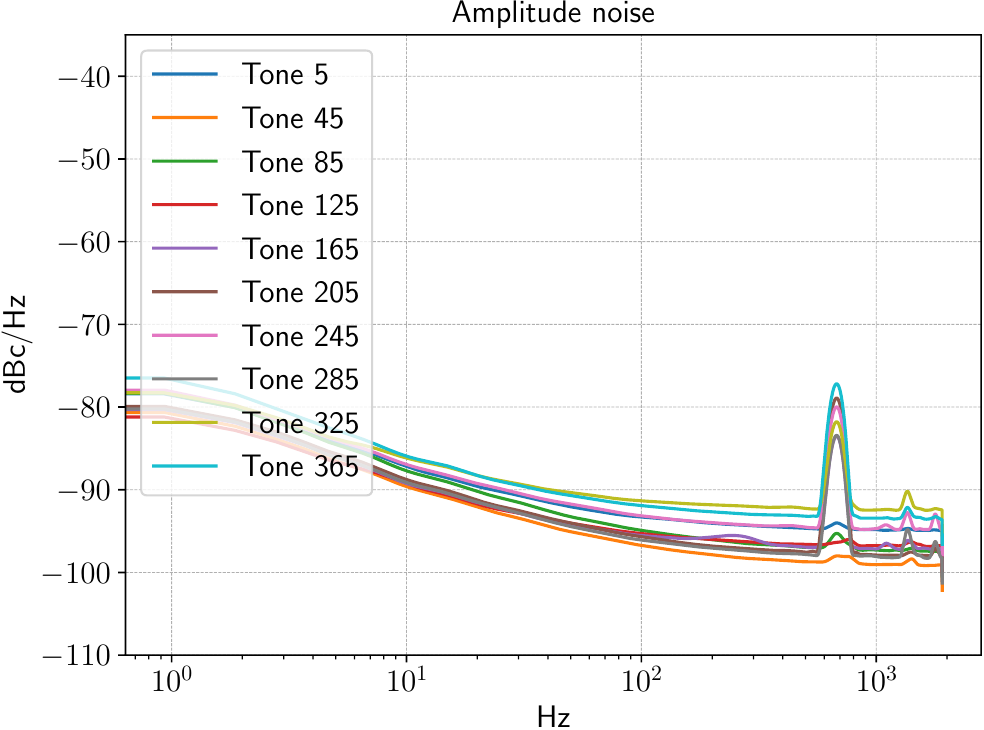}
        \end{subfigure}
        \hfill
        \begin{subfigure}[t]{0.48\textwidth}
            \centering
            \includegraphics[width=\textwidth]{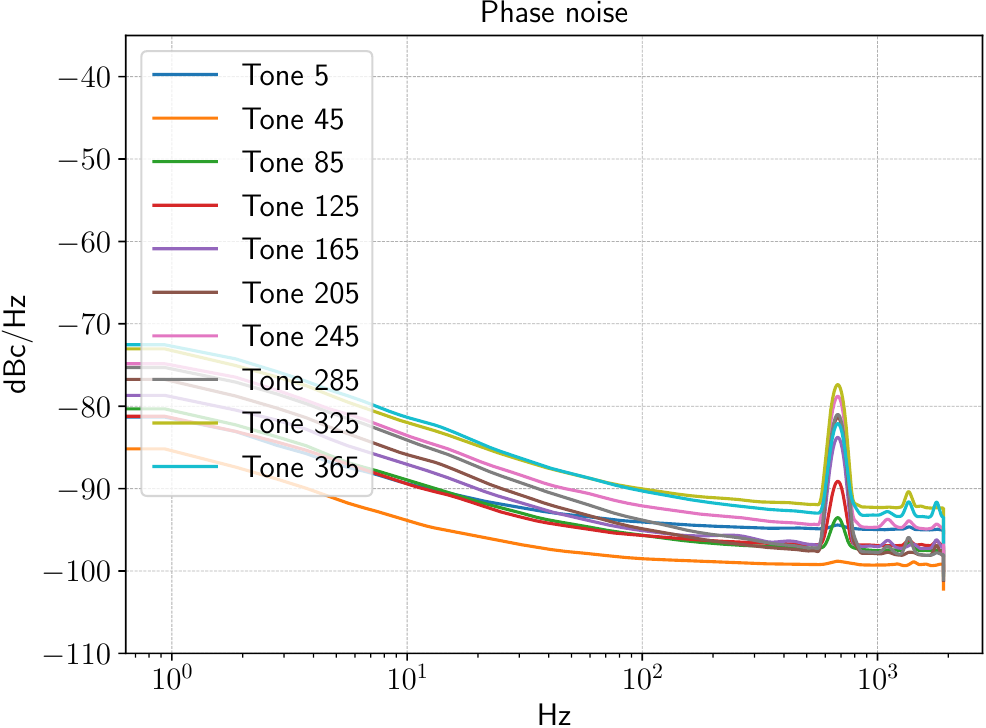}
        \end{subfigure}
        \caption{}
        \label{fig:CORDIC6_MEAS21}
    \end{subfigure}

\caption{Comparison of the measured amplitude and phase noise PSDs using Setup 3~(see Fig.~\ref{fig:setups}), between (a) the original 10-bit, 10-iteration CORDIC firmware, and (b) the proposed modified firmware using a 6-bit, 3-iteration configuration. 
Both implementing the 65,520 samples solution. 
A representative tone from each of the 10~frequency bands is shown.
}
    \label{fig:CORDIC6_MEAS}
\end{figure}

The measurement results, illustrated in Fig.~\ref{fig:CORDIC6_MEAS}, show minimal degradation of the readout performance.  
The tone at 763\,Hz remains at its previous level of approximately \(-80\,\mathrm{dB}\), as shown in Fig.~\ref{fig:CORDIC6_MEAS}\,(b). Meanwhile, the 1526\,Hz tone and the noise floor—the flat plateau following the \(1/f\) noise—exhibit a negligible increase of at most 5\,dB.

\subsection{DDC}

To extract the individual 400~I/Q components of all tones, 400~DDCs are implemented, each requiring two multipliers, resulting in a total of 800~ DSP slices. Considering that only 2,760~DSPs are available, doubling the number of tones would push the design to resource saturation.
To overcome this limitation, we propose a square-wave demodulation technique that preserves the spectral readout performance while requiring no DSP slices. This approach was validated through simulations using the developed twin model, which enabled us to assess the impact of replacing conventional DDC multipliers with DDC square-wave demodulators.

The use of square-wave mixing represents a pragmatic alternative to conventional digital down-conversion due to its low implementation complexity and efficient hardware mapping~\cite{square1, square2, square3}. 

In contrast to traditional DDC architectures that require multiplications with sinusoidal references, square-wave mixing reduces the operation to a binary modulation between \(\pm 1\). This eliminates the need for dedicated multipliers and replaces them with simple sign-control logic, substantially lowering FPGA resource utilization.

In the proposed readout architecture, this concept is further optimized by exploiting signals already available in the system. The sinusoidal waveforms generated by the CORDIC blocks for tone excitation are reused to derive the demodulation references. Specifically, the most significant bit (MSB) of the sine and cosine outputs is extracted, as it directly encodes the instantaneous sign of the waveform. 

The mixing operation is therefore implemented as a conditional sign inversion. During positive half-cycles, the incoming signal is passed unchanged to the averaging filter. During negative half-cycles, its polarity is reversed prior to filtering. This mechanism is mathematically equivalent to demodulating the signal to 0\,Hz by a square wave.

\subsubsection{Signal integrity evaluation}

Using the digital twin, we simulated the entire digital processing chain for both architectures: the original DDC-based implementation and the alternative square-wave demodulation approach. The test signal consists of two superposed tones, $f_1$ and $f_2$, where $f_1$ is the tone of interest. Two stages of the demodulation path are examined: the demodulator output, where $f_1$ is shifted to DC, and the subsequent averaging filter output, where the downconverted signal is low-pass filtered.

With DDC demodulation, four spectral lines are observed: the demodulated tone ($f_1 - f_1 = 0\,\text{Hz}$) and three image components at $f_2 - f_1$, $f_1 + f_1$, and $f_2 + f_1$. Square-wave demodulation produces significantly more frequency components, owing to the rich harmonic structure inherent to square waves: the input signal is mixed not only with the fundamental frequency but also with its odd higher-order harmonics, generating additional spectral replicas.

Despite this increased spectral content, the simulation confirms that all unwanted components are effectively suppressed after the accumulation stage, for both demodulation approaches. This occurs because the spectral images coincide with the zeros of the DDC averaging filter and are therefore rejected. A more detailed analysis of this behavior is provided in previous works~\cite{abdkrimi2025efficient,abdkrimi2025modeling}.

Overall, these results confirm that square-wave demodulation can effectively replace conventional DDC multipliers without compromising signal integrity, a conclusion further validated by the experimental measurements presented in Section~\ref{meas_square}.

\subsubsection{Resource usage}\label{sec:ressourge_usage_DDCQUARE}

In this analysis, we synthesized the complete firmware, as in the previous FPGA resource analysis sections, to compare the resource consumption of the DDC and square-wave demodulation approaches.  
Both implementations rely on the firmware version which incorporates the two key optimizations: the 65,520 samples solution and the 6-bit and 3~iterations CORDIC version.

As shown in Table~\ref{tab:resource_usage_square}, the square-wave demodulation approach significantly reduces DSP usage by 800~DSP blocks (from 1953~to 1153), representing a 28.98\%pt reduction (from 70.76\% to 41.78\%).
However, this comes at the cost of increased logic resource usage:  
LUT consumption increases by 41,929 (from 96,766 to 138,695), a 12.65\%pt increase (from 29.17\% to 41.82\%), and FF usage increases by 2,775 (from 212,218 to 214,993), corresponding to a 0.42\%pt increase (from 31.99\% to 32.41\%).

The observed increase in LUT and FF utilization arises from architectural differences between the two demodulation approaches. In the DDC implementation, two DSP slices are employed, using their embedded multiplier–accumulator structure for demodulation and signal averaging. In contrast, the proposed square-wave demodulation scheme eliminates the use of DSP slices, performing averaging through LUT-based logic instead. 
This design shift inherently increases LUT consumption, while the additional circuitry required for sign inversion based on the MSB introduces further overhead.

Nevertheless, LUT and FF utilization remains well below 50\% of the available FPGA resources, making the trade-off acceptable and not a limiting factor.


\begin{table}[h]
    \centering
    \small
    \renewcommand{\arraystretch}{1.2}
    \caption{FPGA resource utilization for the two demodulation techniques implemented in the readout firmware. 
    Both firmwares use the 65,520 samples solution and the 6-bit, 3-iteration CORDIC configuration}
    
    \begin{tabular}{|c|c|c|c|c|c|c|}
        \hline
        \textbf{Demodulation} & \multicolumn{2}{c|}{\textbf{CLB LUT}} & \multicolumn{2}{c|}{\textbf{CLB FF}} & \multicolumn{2}{c|}{\textbf{DSP}} \\
        \cline{2-7}
        & \textbf{Count} & \textbf{\% Used} & \textbf{Count} & \textbf{\% Used} & \textbf{Count} & \textbf{\% Used} \\
        \hline
        \textbf{DDC} & 96\,766 & 29.17\% & 212\,218 & 31.99\% & 1\,953 & 70.76\% \\
        \hline
        \textbf{Square Wave} & 138\,695 & 41.82\% & 214\,993 & 32.41\% & 1\,153 & 41.78\% \\
        \hline
    \end{tabular}
    
    \label{tab:resource_usage_square}
\end{table}

\subsubsection{Measurement analysis}
\label{meas_square}

To confirm that the proposed approach preserves readout signal quality while significantly reducing DSP resource usage, measurements were performed using Setup~3 (see Fig.~\ref{fig:setups}), which includes all components of KID\_READOUT.  

In both firmware versions, the proposed 65,520 samples solution and the CORDIC configured with 6~bits and 3~iterations were used. 
The goal of this choice is to evaluate whether the performance benefits of these optimizations are maintained when combined with square-wave demodulation.

\begin{figure}[h]
    \centering

    \begin{subfigure}[t]{0.98\textwidth}
        \centering
        \begin{subfigure}[t]{0.48\textwidth}
            \centering
            \includegraphics[width=\textwidth]{figures/CORDIC63AMP-cropped.pdf}
        \end{subfigure}
        \hfill
        \begin{subfigure}[t]{0.48\textwidth}
            \centering
            \includegraphics[width=\textwidth]{figures/CORDIC63PHASE-cropped.pdf}
        \end{subfigure}
        \caption{}
        \label{}
    \end{subfigure}

    \begin{subfigure}[t]{0.98\textwidth}
        \centering
        \begin{subfigure}[t]{0.48\textwidth}
            \centering
            \includegraphics[width=\textwidth]{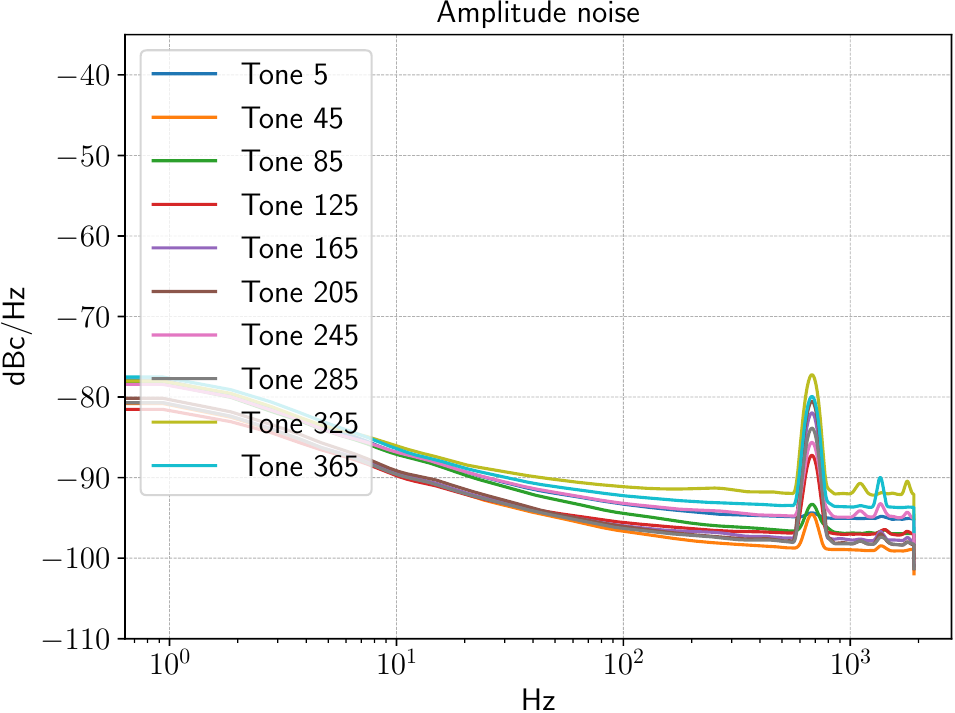}
        \end{subfigure}
        \hfill
        \begin{subfigure}[t]{0.48\textwidth}
            \centering
            \includegraphics[width=\textwidth]{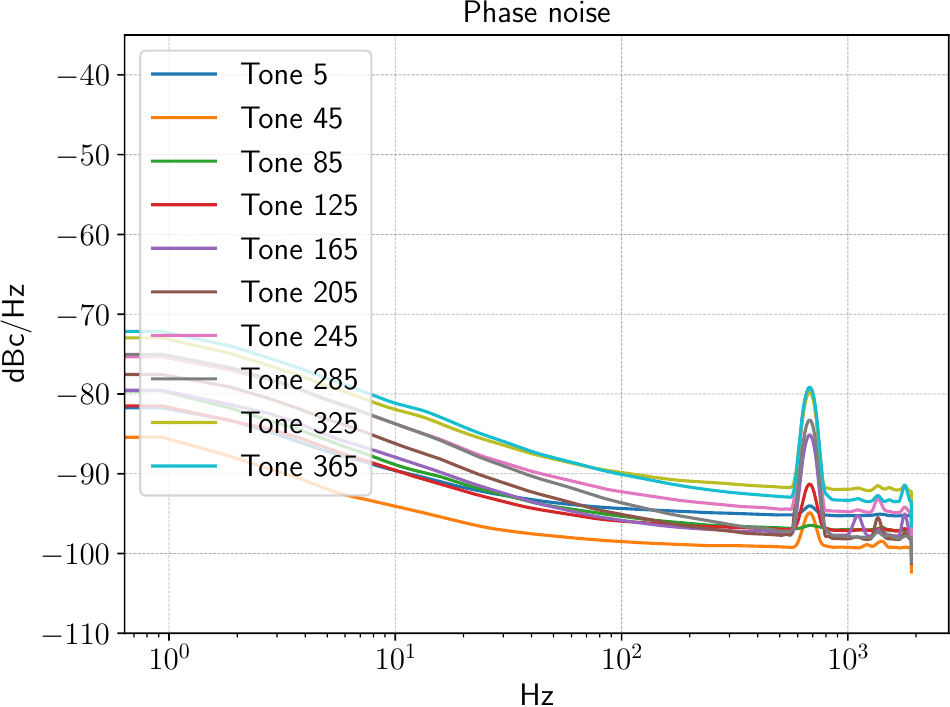}
        \end{subfigure}
        \caption{}
        \label{fig:setup3a}
    \end{subfigure}

\caption{Comparison of the measured amplitude and phase noise PSDs using Setup~3 (see Fig.~\ref{fig:setups}), between (a) the DDC approach and (b) the proposed square-wave demodulation approach.  
Both implementations use the 6 bits, 3 iterations CORDIC configuration and the 65,520 samples solution. 
A representative tone from each of the 10~frequency bands is shown.
}
    \label{fig:Carre_MEAS}
\end{figure}

The measurement results, illustrated in Fig.~\ref{fig:Carre_MEAS}, show identical readout performance.

\section{Instrument-level validation of optimized readout firmware}

This section presents an instrument-level evaluation and validation of the optimized firmware, which incorporates three modifications: 
(1) reduced size of the phase and DDC accumulators from \(2^{16}\) to 65{,}520; 
(2) replaced DDC multipliers with square-wave multipliers; and 
(3) replaced the CORDIC configuration using 10~bits and 10~iterations with the one using 6~bits and 3~iterations.

\subsection{Setup}


The KID\_READOUT board was integrated into the CONCERTO instrument, and the measurements were taken under controlled conditions using a sky simulator, which replicated realistic observational scenarios. 

Prior to data acquisition, a tuning phase was performed to re-identify the resonance frequencies of the MKIDs using a dedicated algorithm~\cite{bounmy2022concerto}.
This step ensured that the excitation tones are aligned with the array frequency resonances.

Following tuning, both firmware versions were deployed under identical experimental conditions (e.g., MKID array, excitation tones, cold electronics, LO). 
Additionally, the acquisition of 655{,}360 I/Q samples per tone was performed as in previous measurements, enabling comparable spectral analysis.

\subsection{Measurements Analysis}
\label{concerto_meas}

Time-domain inspection of the measured I/Q signals revealed several peaks occurring at random intervals, highlighting the instrument's sensitivity to environmental interference. 
To mitigate their impact, we applied an algorithm commonly used in cosmological data analysis. 
This algorithm operates on the time-domain signal by computing its mean \((\mu)\) and standard deviation \((\sigma)\), and then replaces high-amplitude values exceeding \(\mu \pm 5\sigma\) with random values drawn from the interval \([\mu - \sigma,\, \mu + \sigma]\).

\begin{figure}[h]
    \centering

    \begin{subfigure}[t]{0.48\textwidth}
        \centering
        \includegraphics[width=\textwidth]{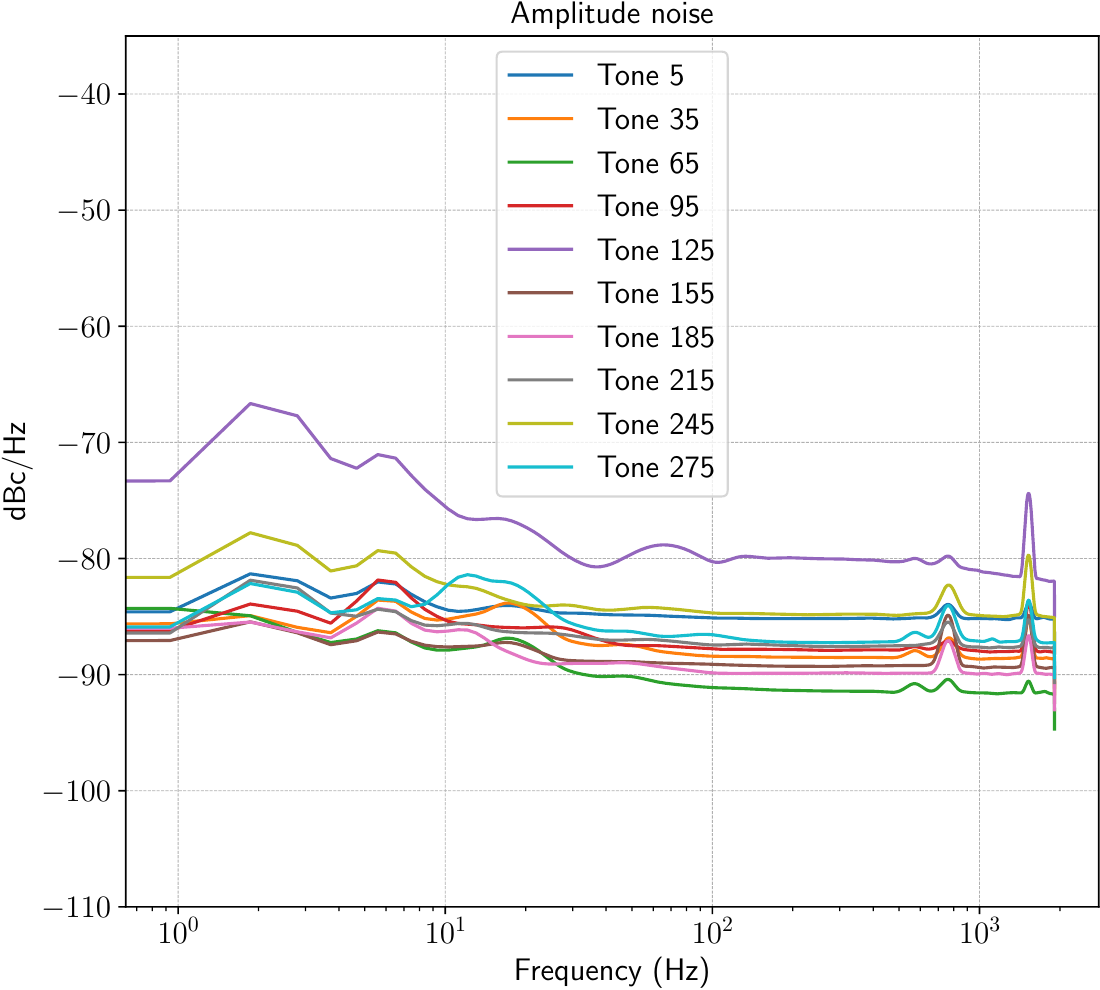}
        \caption{}
        \label{fig:instrum_old_amp}
    \end{subfigure}
    \hfill
    \begin{subfigure}[t]{0.48\textwidth}
        \centering
        \includegraphics[width=\textwidth]{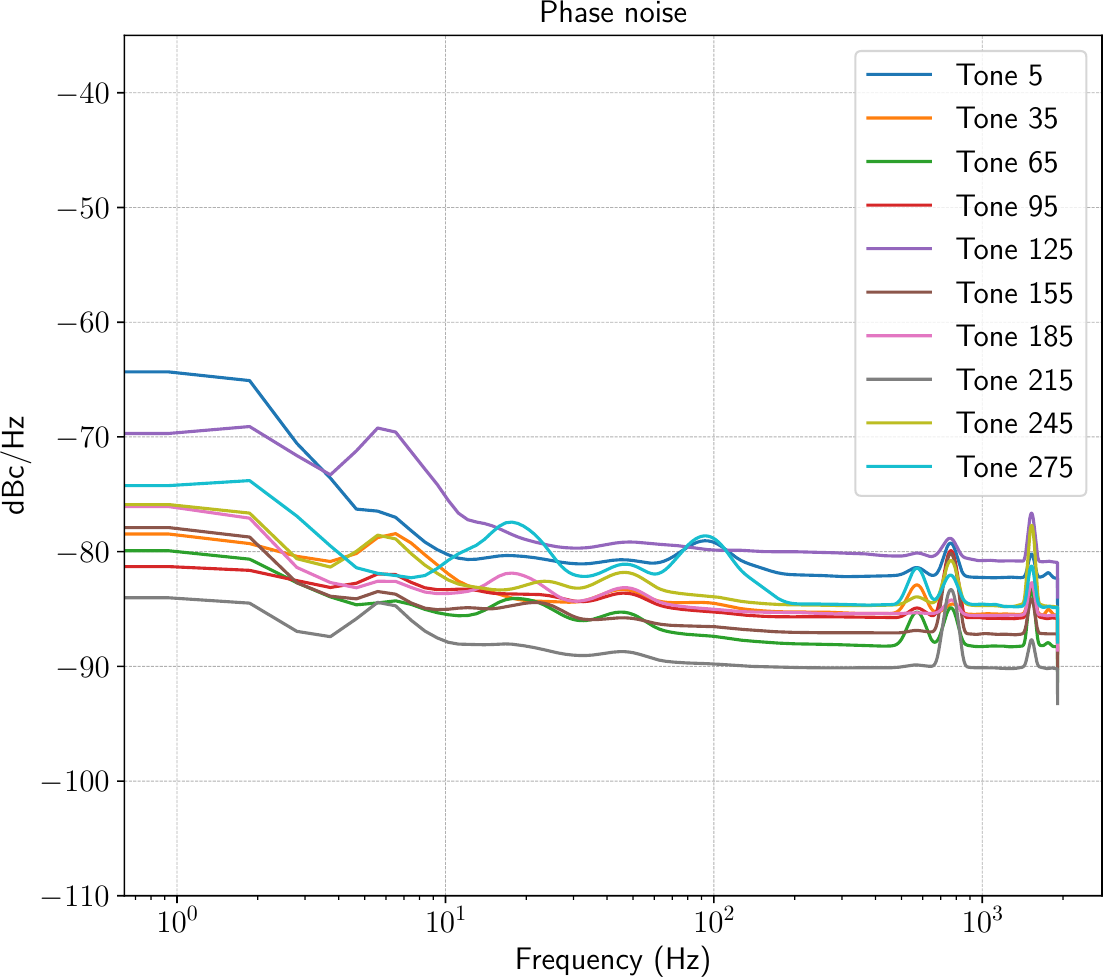}
        \caption{}
        \label{fig:instrum_old_phase}
    \end{subfigure}

    \vspace{0.3cm}

    \begin{subfigure}[t]{0.48\textwidth}
        \centering
        \includegraphics[width=\textwidth]{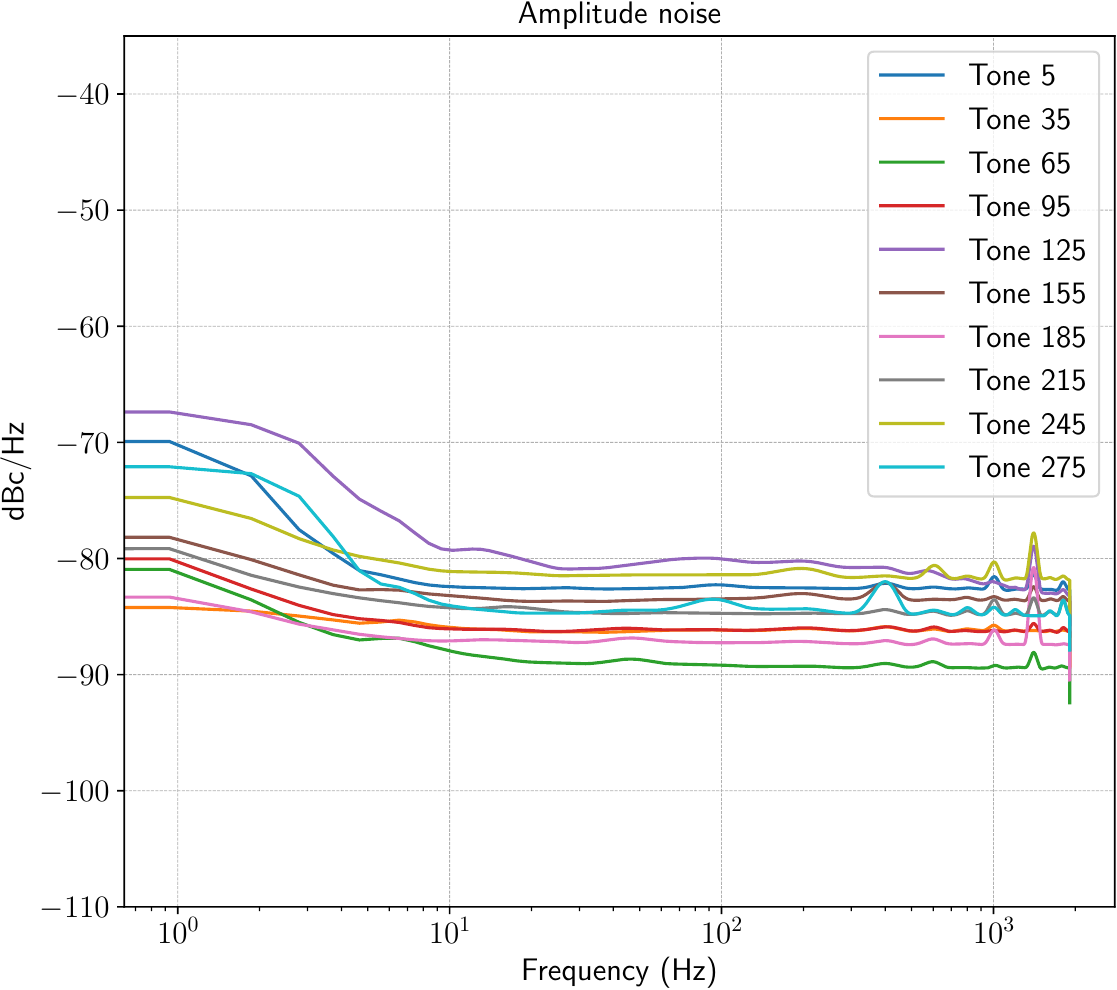}
        \caption{}
        \label{fig:instrum_amp_new}
    \end{subfigure}
    \hfill
    \begin{subfigure}[t]{0.48\textwidth}
        \centering
        \includegraphics[width=\textwidth]{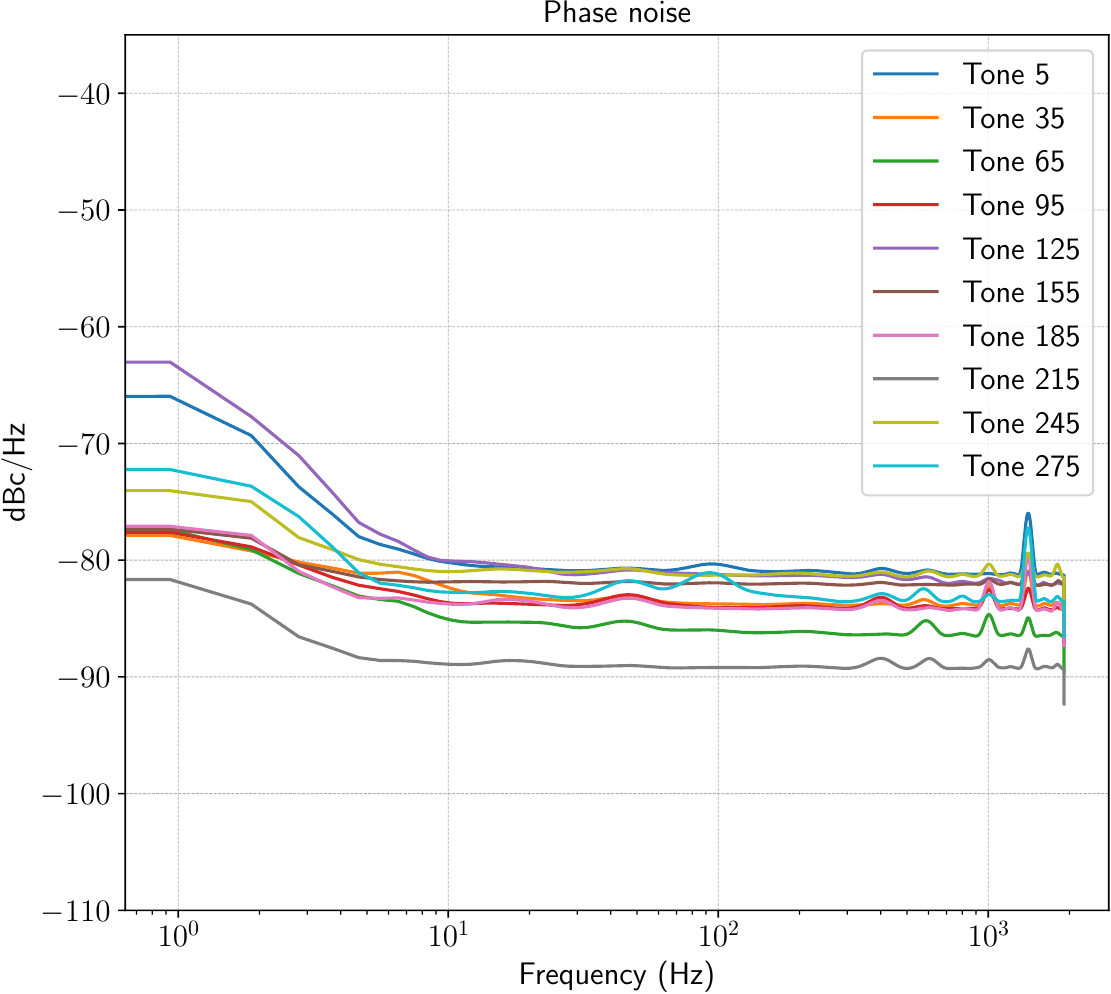}
        \caption{}
        \label{fig:instrum_phase_new}
    \end{subfigure}

    \caption{Comparison of the measured amplitude and phase noise PSDs using the CONCERTO instrument, between the original firmware ((a) and (b)) and the optimized firmware ((c) and (d)).
    A representative tone from each of the 10~frequency bands is shown.}
    \label{fig:instr_valid}
\end{figure}

The measurement spectra shown in Fig.~\ref{fig:instr_valid} exhibit additional spurious components and a higher white noise floor compared to previous measurements performed using the looped-back board alone.

In the standalone electronics measurements, the baseline white noise floor was approximately \(-91\,\text{dBc/Hz}\) for both versions (see Fig.~\ref{fig:setup3_all}(a) and Fig.~\ref{fig:Carre_MEAS}(b)). 
With the original firmware, this level rises to about \(-84\,\text{dBc/Hz}\), and with the new firmware, it increases further to around \(-81\,\text{dBc/Hz}\). 
These elevated noise floors indicate additional noise contributions beyond those present in the KID\_READOUT, originating from the cold electronics or the MKID array itself.
Identifying the dominant source requires further investigation beyond the scope of KID\_READOUT, which was the focus of this work.

Spurs at 763\,Hz and 1526\,Hz, which are visible in the spectra obtained using the original firmware (See Fig.~\ref{fig:instr_valid}(a) and (b)), are no longer present when using the new firmware version (See Fig.~\ref{fig:instr_valid}(c) and (d)).
Their disappearance is attributed to the 65,520 samples solution, which, as previously discussed, effectively attenuates their amplitude.
And because of the elevated white noise, these components are now completely buried and no longer observable.

These results confirm that the new firmware performs reliably and can confidently replace the original implementation for reading out MKID arrays.

\section{Conclusion and perspective}

Simulation of the consolidated digital chain model served as a key diagnostic tool, successfully helping us identify the primary source of the previously unexplained spurs at 763\,Hz and 1526\,Hz, and guided us to implement a mitigation strategy: replacing the original $2^{16}$-sample periodicity with a carefully selected 65,520 samples period for the phase accumulator, used in the tone generator, and the averaging filter of the DDC, used in the tone analyzer.  
Subsequent measurements confirmed the effectiveness of this approach: the spurious tones were entirely eliminated in the fully digital configuration and substantially reduced in setups involving analog components. 
FPGA resource analysis showed that this solution incurred a negligible increase in resource usage—3.24\%pt in LUTs and 0.03\%pt in FFs (See Table~\ref{tab:resource_usage_65520}).

We also optimized two resource-intensive firmware components: the CORDIC and the DDC.  
Through simulation, the 6~bits and 3~iterations CORDIC configuration proved to offer the best trade-off between readout performance and FPGA resource consumption.
Measurements validated this configuration, showing negligible performance degradation while achieving significant savings: a 39.01\%pt reduction in LUT usage and 20.21\%pt in FF consumption (See Table~\ref{tab:resource_usage}).

For the DDC, we introduced an alternative square-wave demodulation technique, eliminating the use of DSP blocks.
Simulations and measurements confirmed that this approach maintains readout signal fidelity.  
As a result, DSP resource usage was reduced by 28.98\%pt. 
The reduced resource load from the CORDIC optimization made room for the increase in LUT and FF usage (12.65\%pt and 0.42\%pt, respectively) caused by the square-wave demodulation approach, while keeping total resource usage well below half of the FPGA limits (See Table~\ref{tab:resource_usage_square}).

Finally, the optimized firmware was implemented and validated through instrument-level measurements.

\begin{table}[h]
\centering
\caption{FPGA resource utilization before and after optimization}
\label{tab:resource_optimization_summary}
\begin{tabular}{lcccc}
\toprule
\textbf{Resource} & \textbf{Available} & \textbf{Original firmware} & \textbf{Optimized firmware} & \textbf{Savings} \\
\midrule
LUTs  & 331,680 & 215,228 (64.89\%) & 138,695 (41.82\%) & $-76{,}533$ (-23.07\%pt) \\
FFs   & 663,360 & 346,539 (52.27\%) & 214,993 (32.41\%) & $-131{,}546$ (-19.86\%pt) \\
DSPs  & 2,760   & 1,953 (70.76\%)   & 1,153 (41.78\%)   & $-800$ (-28.98\%pt) \\
\bottomrule
\end{tabular}
\end{table}

As a result, the same FPGA can now be used to read out twice as many MKIDs per transmission line, fulfilling a central objective of this work and leading to a more efficient design in terms of both resource utilization and overall cost, as summarized in Table~\ref{tab:resource_optimization_summary}.
While the current model is tailored to the existing readout board, its parametric structure allows it to be adapted to future designs, enabling thorough analysis of electronics chain performance to meet the requirements of next-generation MKIDs arrays.

Looking ahead, further efficiency gains could be achieved by replacing the current CORDIC-based sine-wave generation with a square-wave approach in the excitation chain.
The main motivation for this change is hardware efficiency. 
Square-wave generation can be implemented with much simpler logic—relying mainly on counters and comparators to toggle between +1 and –1—resulting in lower resource usage compared to CORDIC.
However, this substitution is not trivial. A rigorous analysis is needed to evaluate its impact on the system. Square waves have higher RMS and peak power than sine waves, which could increase the risk of DAC clipping or exceeding DACs slew-rate~\cite{abdkrimi2024modeling}.
In addition, key performance metrics such as SINAD, SFDR, cross-talk, and hardware resource usage must be carefully studied to determine the best trade-off. 
The model we’ve developed will play a central role in this analysis, helping us assess feasibility and choose the optimal balance between performance and efficiency.

\bibliography{report_sample_bibtex}

\providecommand{\href}[2]{#2}\begingroup\raggedright\begin{thebibliography}{10}

\bibitem{Day2003}
P.~K. Day, H.~G. LeDuc, B.~A. Mazin, A.~Vayonakis and J.~Zmuidzinas, \emph{{A
  broadband superconducting detector suitable for large arrays}},
  {\emph{Nature} {\bfseries 425} (2003) 817}.

\bibitem{klutsch2003modelisation}
I.~Klutsch, \emph{Mod{\'e}lisation des supraconducteurs et mesures}, Ph.D.
  thesis, Institut National Polytechnique de Grenoble-INPG, 2003.

\bibitem{ward2007protostars}
D.~Ward-Thompson, P.~Andr{\'e}, R.~Crutcher and et~al., \emph{Protostars and
  planets v}, {\emph{University of Arizona Press} (2007) }.

\bibitem{bourrion2022concerto}
O.~Bourrion, C.~Hoarau, J.~Bounmy, D.~Tourres, C.~Vescovi, J.-L. Bouly et~al.,
  \emph{Concerto: Readout and control electronics}, {\emph{Journal of
  Instrumentation} {\bfseries 17} (2022) P10047}.

\bibitem{bounmy2022concerto}
J.~Bounmy, C.~Hoarau, J.-F. Mac{\'\i}as-P{\'e}rez, A.~Beelen, A.~Beno{\^\i}t,
  O.~Bourrion et~al., \emph{Concerto: Digital processing for finding and tuning
  lekids}, {\emph{Journal of Instrumentation} {\bfseries 17} (2022) P08037}.

\bibitem{abdkrimi2026spurs}
M.~Abdkrimi, O.~Rossetto, O.~Bourrion, C.~Vescovi and C.~Hoarau, \emph{A
  digital twin of the fpga digital signal processing chain for mkids readout:
  Root-cause analysis and mitigation of spurs}, {\emph{arXiv preprint
  arXiv:2603.04087} (2026) }.

\bibitem{abdkrimi2025cordic}
M.~Abdkrimi, O.~Rossetto, O.~Bourrion, C.~Vescovi and C.~Hoarau,
  \emph{Optimized fpga implementation of the cordic algorithm for a frequency
  multiplexed readout},  in \emph{2025 14th Mediterranean Conference on
  Embedded Computing (MECO)}, IEEE, 2025.

\bibitem{square1}
D.-H. Ha and B.-J. Kim, \emph{Low-power quadrature demodulation using
  square-wave mixers in cmos rf receivers}, {\emph{IEEE Transactions on
  Circuits and Systems II: Express Briefs} {\bfseries 66} (2019) 2042}.

\bibitem{square2}
J.~Alvarez, P.~Garcia and A.~Garcia, \emph{Fpga implementation of a digital
  lock-in amplifier based on square wave demodulation},  in \emph{2020 IEEE
  International Symposium on Circuits and Systems (ISCAS)}, pp.~1--5, 2020.

\bibitem{square3}
S.~Chattopadhyay and K.~Mandal, \emph{Real-time spectral analysis using
  square-wave demodulation for software defined radios}, {\emph{Journal of
  Signal Processing Systems} {\bfseries 94} (2022) 501}.

\bibitem{abdkrimi2025efficient}
M.~Abdkrimi, O.~Rossetto, O.~Bourrion, C.~Vescovi and C.~Hoarau,
  \emph{Efficient fpga readout architecture for mkids: A dsp-light approach},
  in \emph{2025 20th International Conference on PhD Research in
  Microelectronics and Electronics (PRIME)}, pp.~1--4, IEEE, 2025.

\bibitem{abdkrimi2025modeling}
M.~Abdkrimi, \emph{Modeling of the readout chain and optimization of digital
  signal processing on FPGA for superconducting kinetic inductance detectors},
  Ph.D. thesis, Universit{\'e} Grenoble Alpes [2020-....], 2025.

\bibitem{abdkrimi2024modeling}
M.~Abdkrimi, O.~Rossetto, O.~Bourrion, C.~Vescovi and C.~Hoarau, \emph{Modeling
  and analysis of digital-to-analog converter non-idealities in microwave
  kinetic inductance detectors readout},  in \emph{2024 IEEE 28th Workshop on
  Signal and Power Integrity (SPI)}, pp.~1--4, IEEE, 2024.

\end{thebibliography}\endgroup
\bibliographystyle{JHEP}

\end{document}